\documentstyle[aaspptwo]{article}
\newcommand{\etal}{{{ et al.}}~}
\newcommand{\eg}{{{e.g.,}}~}
\newcommand{\ie}{{{i.e.,}}~}
\newcommand{\kms}{{{km s$^{-1}$}}~}
\newcommand{\kmsc}{{{km s$^{-1},$}}~}
\newcommand{\kmsp}{{{km s$^{-1}.$}}~}
\newcommand{\Ho}{{{H$_\circ$}}~}
\newcommand{\amin}{{{$^\prime$}}~}
\newcommand{\asec}{{$^{\prime\prime}$}~}

\begin{document}

\title{ Anatomy of a Merger: A Numerical Model of A754}

\vspace{1.in}
\author{KURT ROETTIGER$^1$}
\affil{Goddard Space Flight Center\\Code 930\\Greenbelt, MD, 20771\\e-mail: kroettig@pecos1.gsfc.nasa.gov}
\author{JAMES M. STONE}
\affil{Department of Astronomy \\ University of Maryland\\ College Park, MD 20742-2421
\\jstone@astro.umd.edu}
\author{RICHARD F. MUSHOTZKY}
\affil{Goddard Space Flight Center\\Code 662\\Greenbelt, Maryland 20771\\
e-mail: mushotzky@lheavx.gsfc.nasa.gov}

\vspace{.5in}



\altaffiltext{1}{NAS/NRC Associate}


\begin{abstract}
A754 is a well-observed cluster of galaxies (z=0.054) which exhibits a variety
of morphological peculiarities. These include a bar of X-ray emission that
is offset significantly from the galaxy distribution, an elongated
X-ray surface brightness distribution extending between two distinct
peaks in the galaxy distribution, and an extremely non-isothermal
and asymmetric intracluster medium (ICM) temperature morphology. Using these
observational constraints, we present a numerical
Hydro/N-body model of A754 in which two clusters (2.5:1 mass
ratio) have merged nearly in the plane of the sky less than 0.5 Gyrs
ago with an impact parameter of $\sim$120 kpc, and an 
impact velocity of $\sim$2500 \kms (roughly the escape velocity of the
primary cluster).  Our models allow us to identify the origin of A754's peculiar X-ray and
temperature morphologies, the underlying hydrodynamical processes
that shape them, and their future evolution.  We make detailed predictions for future high
resolution X-ray spectroscopic observations (e.g. ASTRO-E).  We discuss
general properties of our models which will be characteristic of off-axis mergers.
In particular, we find significant non-thermal pressure support within the central
region which could bias cluster mass estimates. We find significant angular momentum
imparted on the gas distribution in the cluster. We find that mixing of the subcluster
gas components is an inefficient process, particularly at large radii. Finally,
we find that subsequent dark matter core passages result in an extended relaxation
timescale.

\end{abstract}


\section { INTRODUCTION}

In recent years, clusters of galaxies have become essential cosmological probes. Both their
current dynamical state and internal structure provide important clues to conditions in
the early universe. Over the years,  many researchers
(\eg Geller \& Beers 1982; Dressler \& Shectman 1988; West \& Bothun 1990; Jones \& Forman 1991; 
Bird 1994; Davis 1994; Mohr \etal 1995, among others) using both X-ray and optical
databases, have attempted to understand the dynamical state of clusters through analysis of 
their projected substructure (\ie clumpiness and asymmetries in the distribution of the intracluster medium (ICM)
and galaxies). Unfortunately, these studies are inherently limited because they 
represent only a 2-dimensional projection of the true 3-dimensional cluster shape, and moreover,
there is no information regarding the gas kinematics. As such, the current data set cannot
reveal the underlying physical processes
which produce the observed substructure nor can they reveal 
how the substructure will evolve in time.  For these answers, one must turn to numerical simulations
(\eg Evrard 1990; Cen 1992; Roettiger \etal 1993; Schindler \& M\"uller 1993; Pearce, Thomas \& Couchman 1994; 
Navarro \& White 1994; Mohr \etal 1995, among others). In this paper, we apply numerical
hydrodynamical/N-body simulations to the analysis of the extensive observational database collected on the
galaxy cluster Abell 754.

 A754, at z=0.054, is one of the most extensively observed clusters in both the optical and X-ray bands.
Much attention has been drawn to this particular cluster by its peculiar
 X-ray morphology (Fabricant \etal 1986, F86; Henry \& Briel 1995, HB95). 
Far from being a relaxed spherical 
distribution centered on the galaxies, A754 shows an elongated and barred X-ray morphology significantly
offset from a bimodal galaxy distribution. Even more recently, A754 has been recognized
for its significant non-isothermality (HB95; Henriksen \& Markevitch 1996, hereafter HM96). Together, 
these properties make A754 an
ideal laboratory for the study of non-equilibrium systems.

The purpose of this paper is to assemble the observational database on
A754 and  create a plausible, though not necessarily unique, self-consistent, numerical model
 that is in general agreement with the observational data. The model is then used
to explore the underlying hydrodynamical processes that produced the observed
morphological features, with particular emphasis on the X-ray surface brightness
and temperature distribution. We then follow the development of these features
as the merger is allowed to evolve thus giving insight into the more general evolution
of a slightly offaxis merger.

We perform this study using a hydrodynamical/N-body code based on the piecewise parabolic
method (PPM) and a particle-mesh algorithm (PM). The gaseous intracluster
medium (ICM, as represented by PPM) is allowed to evolve self-consistently (including self-gravity)
within a changing gravitational potential defined largely by the dark matter distribution
as represented by the collisionless N-body particles. In this study, we employ idealized
initial conditions in a manner similar to previous studies by Roettiger \etal (1993, 1995, 1996, 1997a),
 Pearce \etal (1994), and Ricker (1997). The Pearce \etal simulations are based on a Smoothed Particle Hydrodynamics
code (SPH, Lucy 1977), while Ricker (1997) explores more extremely offaxis mergers.
The idealized initial conditions used here allow us to explicitly control the merger parameters making possible a systematic
survey of parameter space. It also allows for an efficient use of the computational volume
providing improved resolution over simulations of largescale structure formation.
On the other hand, our study is limited by  the neglect of largescale tidal forces,
cosmological infall, and  uncertainty in the detailed structure of clusters.
The cosmological infall of baryonic matter is not likely to affect the inner regions
of the cluster which are of interest here. Of potentially greater importance, these simulations
do not include radiative cooling which may influence the cluster's thermal evolution,
particularly in the high gas density core.

In \S \ref{data} we review the recent optical and X-ray observations
of A754. Section \ref{num} describes the numerical method, the code, grid configuration, etc. In 
\S \ref{constrain}, we discuss the degrees of freedom in
the model and the manner in which the observations are used to constrain the
model parameters. Here, we also describe the initial conditions and summarize the
merger parameters. In \S \ref{comp}, we make a quantitative comparison of our model with the
observational data while discussing the current dynamical state of A754. Section \ref{evol} describes
the future evolution of the system. We summarize our results in \S \ref{summary}
We define $h=$\Ho$/100$ for consistency with previous  observational results,
but we choose \Ho=65 \kms Mpc$^{-1}$ for the scaling of our model.

\section {THE OBSERVATIONAL DATA}
\label{data}
\subsection {Optical Data}
\label{optical}

Many studies have been performed on the distribution of galaxies
in A754. Of these, many have found significant substructure (\eg Geller \& Beers 1982;
F86; Escalera \& Mazure 1992 (EM92); Bird 1994; and Zabludoff \& Zaritsky 1995 (ZZ95))  while
others have not (\eg Dressler \& Shectman 1988, West \& Bothun 1990). 
Of those studies that do find substructure, two
primary galaxy concentrations are consistently identified. Their spatial
location relative to the X-ray surface brightness distribution is
indicated in Figure \ref{obsdata} (see also Plate L8 in ZZ95). Hereafter, we refer
to these as the SE and NW concentrations. Table 1 is 
a summary of the individual subcluster properties as determined by 
several researchers.

Examination of Table 1 reveals a general consistency between the various
analyses. The largest discrepancy  resides in the velocity dispersion
of the NW population determined by EM92. Their value differs from the other
two by greater than 2$\sigma$ and is less than the dispersion calculated
for the SE group. In each of the other two studies, the NW group has a larger
velocity dispersion than the SE group. It is also interesting to note
that in each study more galaxies are found to be associated with the
SE subcluster than with the NW subcluster. 

The various estimates of A754's virial mass are consistent although
uncertainties are large. Biviano \etal (1993) find
4.07$\pm^{5.05}_{2.23}$ $\times$ 10$^{14}$  M$_\odot$ and 6.03$\pm^{4.67}_{2.65}$ 
$\times$ 10$^{14}$  M$_\odot$ within 0.75$h^{-1}$ and 1.5$h^{-1}$ Mpc,
respectively. Escalera \etal (1994) find 10.48$\pm3.27$ $\times$ 10$^{14}$ M$_\odot$ corresponding to a virial
radius of 2.21$\pm0.28$$h^{-1}$ Mpc. Only EM92 have attempted to determine
masses for the individual concentrations. They find 2.9 $\times$ 10$^{14}$ M$_\odot$ and 1.3 $\times$
10$^{14}$  M$_\odot$ within $\sim$0.32 Mpc for the SE and NW concentrations,
respectively.

To summarize, the data are consistent with two clusters
at nearly identical redshifts having a projected separation of
0.73$h^{-1}$ Mpc (ZZ95). Their velocity dispersions are
of order 900 \kms with the ratio of their dispersions consistent
with unity. The total mass of the system within $\sim$2$h^{-1}$ Mpc, based
on the galaxy dynamics, is likely greater than 10$^{15}$  M$_\odot$.

\subsection  { X-ray Data}
\label{xray}

The ROSAT X-ray surface brightness image generated by HB95
reveals an unusually complex morphology. We have reproduced this data in 
Figure \ref{obsdata} (solid contour).
There are several features worth noting. First, the general elongation which
extends parallel to the line connecting the galaxy concentrations (\S \ref{optical}).
Second, the peak of the X-ray distribution is seen to be a well-defined bar which 
extends up to $\sim$4\amin to the north of the SE galaxy concentration and is nearly
perpendicular to the line connecting the galaxy concentrations. In a wavelet
analysis of the X-ray surface brightness distribution (Slezak, Durret, \& Gerbal 1994,
Figure 16 therein.), the peak of the distribution is associated with an angled
 feature which includes the bar and a less well-defined extension toward
 the NW galaxy concentration. The total effect is that the eastern edge of the bar
is sharper than the western edge.  Finally, it is important to note that
neither of the major galaxy concentrations are spatially coincident with the
peak X-ray emission (ZZ95). The dominant galaxy in A754, identified as a cD galaxy
(Dressler 1980), is located in the NW subcluster, 13\amin from the X-ray
peak.

The mean temperature of the X-ray emitting gas in A754 is 9 keV (HM96). Using the
global velocity dispersion of 750 \kms (Mazure \etal 1996), $\beta=\mu m_h \sigma_v^2/k T$ is found
to be 0.4 which is
significantly discrepant from the best-fit mean value of 0.94$\pm0.08$ found by
Lubin \& Bahcall (1993) for a large sample of nearby clusters.  In any case,
the mean temperature may be misleading in that A754 is decidedly non-isothermal.
Henriksen (1993), using HEAO-1 A2 data, identified two ill-defined temperature 
components within A754. HB95, using ROSAT data, were
the first to produce a detailed temperature map of A754. They found a relatively 
cool region near the X-ray peak at $\sim$5-7 keV. To the north and south
they identify arcs of gas greater than 9 and 10.8 keV, respectively.
To the west, they identify an arc with temperature greater than 13.8 keV.

 A similar attempt to map the temperature distribution was made by HM96 using ASCA data. 
Although differing in detail, there is considerable consistency between the HB95 and HM96
maps. Only one of the nine central regions (as defined by HM96, Figure 1, therein)
differed by a statistically significant amount from the corresponding
region in HB95. Even in the low surface brightness outer regions
of the cluster where the detailed energy dependence of the ASCA point spread function and
the background subtraction are both important, there is considerable agreement between
the two maps. It is likely that much
of the observed discrepancy results from the different energy bands employed. The low
energy ROSAT band (0.1-2.5 keV) is less able to constrain the temperatures of the extremely
hot gas observed by ASCA which is sensitive to the 0.5-10 keV band. 

Finally, for completeness, we comment on the evidence for a cooling flow in A754.
Edge, Stewart \& Fabian (1992) report evidence for a moderate cooling flow, $\sim$24 M$_\odot$/yr.
However, there is no evidence for the optical filaments often associated with 
cooling flows (Heckman \etal 1989). Both HB95 and HM96 report relatively low temperatures ($\sim$6 keV)
near the X-ray maximum. It as been suggested that this is cool gas having been
stripped from one of the subclusters (HM96). It is also possible that this is a
remnant of a cooling flow that was disrupted by the merger.  No attempt was made to 
model a premerger cooling flow
although we would expect one to have been disrupted during the merger (G\'omez \etal 1997).
If there had been a long-established cooling flow in A754 before the merger, it is quite likely
that remnant cool gas would remain at this early epoch of the merger owing to the
inefficient mixing of cluster gas components (See \S \ref{mixing}).

\section{NUMERICAL METHOD}
\label{num}
We compute the dynamical evolution of merging clusters of galaxies using
 a hybrid hydrodynamical/N-body code in which the hydrodynamical component is
CMHOG written by one of us, J. M. Stone.  CMHOG solves the fluid equations using
 an implementation of the piecewise-parabolic method 
(PPM, Colella \& Woodward 1984) in its Lagrangian remap formulation: its fidelity
has been demonstrated through numerous applications in interstellar gas dynamics,
\eg the interaction of strong shocks with dense clouds (Stone \& Norman 1992).
The collisionless dark matter is evolved using an N-body code based on a standard 
particle-mesh algorithm (PM, Hockney \& Eastwood 1988). The particles
are evolved on the same grid as the gas using the same time step. The time step
is determined by applying the Courant condition simultaneously to both the dark 
matter and the hydrodynamics.
The only interaction between the collisionless particles and the gas is 
gravitational. Since we are modeling an isolated
region, the boundary conditions for Poisson's equation are determined by a multipole expansion
of the mass distribution contained within the grid (Jackson 1975).  Particles that leave the grid are
lost to the simulation. Typically, less than a few percent of the particles leave the grid.
Previously, this code has been used to examine systematic errors in the measurement of the
Hubble constant using Sunyaev-Zeldovich effect in merging clusters of galaxies (Roettiger,
Stone, \& Mushotzky 1997).

The simulation is fully three-dimensional. Two different computational grids were used. A
survey of parameter space was performed on the MasPar-2 at Goddard Space Flight Center (GSFC) using a fixed 
and rectangular grid
(256 x 128 x 128 zones) having linear dimensions of 12.8 x 8.3 x 8.3 Mpc. 
The merger axis coincides with the grid's major axis along which 
 resolution is uniform and scales to 50 kpc or $\sim$5 zones per
primary cluster core radius. Resolution along the grid's minor axis is uniform within 
the central 
64 zones and ratioed in the 32 zones on either side. That is, the resolution
is a uniform 50 kpc extending 1.6 Mpc (32 zones) on either side of the major axis. Beyond 1.6 Mpc,
the zone dimensions increase by $\sim$3\% from one zone to the next out to the edge of the grid.
We use outflow boundary conditions for the hydrodynamical evolution. Once we
settled on a set of initial conditions, we ran a very high resolution simulation on the 
CM-5 at the National Center for Supercomputing Applications (NCSA) using a fixed rectangular 
grid  with dimensions of 512 x 256 x 256 giving a resolution of 25 kpc in the central 3.2 Mpc (128 zones),
or nearly 10 zones per core radius.

\section {MODEL CONSTRAINTS}
\label{constrain}

\subsection{The Data and its Limitations}

We use the above observations  to constrain our numerical model. However, there are
serious limitations to this process. First, the observational
uncertainties can be large.  Second, we believe, as did ZZ95, that this is
a major merger in the very early stages after core passage. If this is the case,
then the structure, temperature and even the mass of the initial systems can be
severely obscured by the current non-equilibrium conditions (Roettiger \etal 1996).
 Finally, there are just too many free
and not necessarily independent parameters to completely constrain the model. 
These include, the total mass of the system, the relative masses of the individual
subclusters, the distribution of mass within the subclusters ($\rho \sim r^{-n}$, core
sizes, etc), the fraction of mass contained in baryons, the distribution
of baryons relative to the dark matter (\ie the $\beta$ parameter, central gas
density), the angular momentum of the system (impact parameter and velocity),
the linear scaling (we assume \Ho=65 \kms Mpc$^{-1}$),
the merger epoch (time since closest approach), and the orientation of the merger
with respect to the observer.  For these reasons, it is difficult to
claim a definitive model. Rather, we have produced a model which
agrees well enough with the observational data that we may consider it to be a
plausible representation of the gasdynamics within A754.

Our approach is to use the observable properties of the general cluster population
to construct our initial clusters. We then explore essentially two
branches  of parameter space (relative central gas density and impact parameter, \S\ref{param}) at 
low resolution (50 kpc per grid zone), determine a best fit, then run a high resolution
 (25 kpc per grid zone) simulation which is discussed in \S \ref{comp}

\subsection{Initial Conditions}
\label{init}
Since it is likely that much of the information about the internal structure of
the premerger clusters
has been erased by the merger, we begin with clusters that are consistent
with observations of presumably relaxed systems. Both subclusters were chosen
to have the same mass distribution. For our clusters, we have chosen the
lowered isothermal King Model described in Binney and Tremaine (1987). The
lowered isothermal King Model is a family of mass distributions characterized
 by the quantity $\psi/\sigma^2$ which essentially defines
the concentration of matter. As $\psi/\sigma^2$ increases, the concentration
parameter also increases, \ie the core radius ($r_c$) decreases relative to the 
tidal radius, $r_t$.
We have chosen a model with  $\psi/\sigma^2$=12 in which we have  truncated the
density distribution at 15$r_c$. Near the half mass radius, the total mass density
follows a power law distribution, $\rho=r^{-\alpha}$ where $\alpha \sim$ 2.6.
This model is consistent with mass distributions produced by the cosmological
N-body simulations of Crone \etal (1994) which showed $\alpha \sim$ 2.4 in a high
density universe ($\Omega$=1) and $\alpha \sim$ 2.9 in a low density universe
($\Omega$=0.2). Observationally, galaxies are distributed as $\alpha \sim$ 2.4$\pm$0.2
(Bahcall \& Lubin 1994).

The simulations are conducted in scaled units. They are rescaled to
meaningful physical dimensions after the fact. This gives us some flexibility
in defining the initial conditions in that dimensionless
subcluster parameters are more important than the global scaling. The
observational data does give some insight into the scaling of global parameters.
The various mass estimates listed in \S \ref{optical} seem to indicate that the
total mass within 2-3 Mpc is likely greater than 1.0 $\times 10^{15}$ M$_\odot$.
The observed redshift of A754 indicates that the galaxy concentrations
are separated by $\sim$1.12 Mpc for \Ho=65 km s$^{-1}$ Mpc$^{-1}$. 

The attempts to separate the two galaxy components can be used to constrain
their relative masses. There is considerable uncertainty in doing so
since at this stage of the merger the velocity dispersions may not
be representative of the individual masses. We have found in
previous (\eg Roettiger \etal 1993) simulations that subclusters are ``heated" relative to the
primary as they pass through or near its core. Consequently, they may appear more
 massive than they actually are. Also, as seen in Table 1,
small number statistics lead to large uncertainties in the velocity dispersions. The ratio of the
NW to SE subcluster velocity dispersions are consistent with unity which would
indicate that they are of comparable mass (assuming heating is not significant).
Taken at face value, the velocity dispersions in F86 and ZZ95 indicate that the NW subcluster is somewhat
more massive than the SE subcluster, while EM92 indicates that the SE subcluster
is more massive.
Looking at the galaxy counts in Table 1 and assuming that the galaxies
trace the dark matter mass and that significant exchange of galaxies has not occurred,
all three samples in Table 1 indicate  that the SE subcluster is the more 
massive subcluster. HB95 also suggests that the SE subcluster is the original
main cluster based on its central location relative to the presumably undisturbed
outer X-ray contours. For these reasons, we consider mergers where 
 the SE subcluster is the more massive cluster by a factor of 2.5. Although marginally consistent with
the velocity dispersions of F86 and ZZ96, it does represent one extreme, and a 
lower mass ratio (approaching 1:1) cannot be ruled out.

\subsection{Parameter Space}
\label{param}

Having settled on the relative cluster masses, we then explored a range of
impact parameters and relative central gas densities.
We examined impact parameters ranging from zero (head-on) to the sum of the respective core radii (Table 2).
We found that in order to generate the combination of compression and shear necessary to
produce the X-ray core morphology, the cores of the interacting clusters must
overlap. Similarly, for a given total mass ratio, the relative ram pressure
experienced by the clusters is determined by the relative central gas densities.
The central gas density ratio ($n_{eo1}/n_{eo2}$) was varied from 0.5 to 2.0. The goal
was to have enough ram pressure to create the desired X-ray core morphology
but not so much that the subcluster penetrates the primary core. 

The central gas densities are determined by a combination of the global
gas fraction (f$_b$) and the distribution of the gas relative to the 
dark matter (\ie the $\beta$-parameter). We do not attempt to determine
the global gas fraction from the data. Rather, we arbitrarily choose the
more massive cluster to have f$_b$=0.12 and $\beta$=0.75 (See \S\ref{xray}). Both values are
within the acceptable range defined by observation (\eg Jones \& Forman 1984; White \& Fabian 1995;
Lubin \& Bahcall 1993). We then varied these parameters for the
less massive cluster, also within the observed range, in order to obtain
the desired central gas density ratio.

Below we briefly summarize the initial cluster parameters. We further
discuss the specific effects of parameter space in \S\ref{comp}

\subsection{A Brief Model Summary}

Table 2 contains the initial cluster parameters. These clusters
were allowed to merge under the influence of their mutual gravity.
They are given an initial velocity of 270 \kms parallel to the line of
centers and 100 \kms perpendicular. This results in a slightly off-axis
merger with an impact parameter of $\sim$120 kpc and a final impact
velocity of 2700 \kmsp Both of these values are relative to the respective
centers of mass of each cluster.  The simulation most closely resembles
A754 at $\sim$0.3 Gyrs after closest approach. Although the
relative masses of the two clusters is poorly constrained, it is
important that the SE component (the primary cluster in our model) have
the greater central gas density initially. It is also important
that the NW component (secondary cluster in our model) have a relatively
more concentrated mass distribution.  
Below, we discuss the justification for the choice of these parameters.

\section{A COMPARISON WITH THE DATA}
\label{comp}

Figure \ref{simdat1} contains plots of four quantities overlaid
with the X-ray surface brightness contours. They include (a) the projected,
emission-weighted ICM temperature, (b) the dark matter distribution,
(c) the gas velocity in the plane of the merger, and (d) a dynamically inert 
quantity used to trace the two gas components, also within a slice
taken in the plane of the merger. In each case, the data is taken from 
the high resolution simulation (see \S \ref{num}), but it represents
only a small fraction of the total volume (80 x 80 zones, in projection). 
The X-ray emissivity was calculated using the Mewe-Kaastra-Liedahl emission
spectrum for optically-thin plasmas supplied with XSPEC (Arnaud 1996). The X-ray surface
brightness is simply a line-of-sight integration of the volume emissivity.
Similarly, the emission-weighted, projected ICM temperature is simply a line-of-sight
integration of the product of the temperature within a given zone and the X-ray emissivity in
that same zone divided by the total emissivity along a given line-of-sight.
The dark matter distribution is the projected particle density. The merger
is assumed to be exactly in the plane of the sky.
The epoch depicted is 0.3 Gyrs after
the closest approach of the respective dark matter centers of mass.
The smaller of the two clusters has moved left to right passing
below the core of the more massive cluster. Hereafter, we refer to
the lower left dark matter concentration as the primary
cluster, and to the upper right dark matter concentration  as
the subcluster. When comparing to the observational data, lower left is SE, upper
right is NW.  Table 3 shows a comparison of some of the global cluster
parameters.

\subsection{X-ray Morphology and Hydrodynamics}
\label{xmorph}

The simulated X-ray surface brightness morphology at 0.3 Gyrs is
represented by contours in each of the four panels in Figure \ref{simdat1}.
Using our model, we are able to reproduce the basic features
of A754's X-ray morphology (\S\ref{xray}, Figure \ref{obsdata}). These include, the bar-shaped peak
in the X-ray emission, the displacement of the X-ray peak relative
to the galaxies/dark matter, and the overall east-west elongation of the
X-ray emission. The relationship between the orientation of the emission peak to
the overall X-ray surface brightness distribution represent an isophotal twisting
of nearly 90 degrees. We now identify the underlying processes responsible
for the morphological features exhibited by the simulated data.

The flattening of the X-ray peak on the eastern
side is a result of the extreme ram pressure ($\rho v^2$) of the subcluster
on the  primary.  At the
time of closest approach (as defined by the
respective dark matter centers of mass) the relative velocity is $\sim$2500 \kmsp
 Since the merger
is slightly off-axis, a portion of the subcluster (moving east to west)
essentially runs into a wall of gas (\ie the core of the primary cluster)
with sufficient ram pressure to flatten the distribution and displace
it relative to the dark matter. Figure \ref{simdat1}c shows the basic flow
pattern within a 2-D slice in the plane of the merger. The maximum gas velocities
at this epoch are associated with the residual infalling subcluster
gas located SE of the X-ray peak. These gas velocities are seen to approach 1800 \kms
(relative to the primary
cluster dark matter distribution), but they abruptly decrease in magnitude ($<$1000 \kms) 
as they pass through a shock to the SE of the core after which 
the infalling gas is deflected by the core toward the SW and NE. Still, there is
sufficient ram pressure to displace the X-ray peak from the gravitational
potential minimum by a significant distance. 

ZZ95 finds the peak of the X-ray emission to be displaced by 3.8\amin from
the peak of the SE galaxy concentration (assumed to be the minimum of the
gravitational potential). Looking at the more statistically
robust centroids of these distributions, the offset is somewhat less, 
$\sim$3.0\amin.
Our model shows a dark matter/X-ray offset of 100 kpc which is $\sim$1.5\amin assuming
\Ho=65 \kms Mpc$^{-1}$. Although somewhat less than the observed value, the simulated
offset is in the same direction as that seen in the data and of comparable magnitude.  A larger offset
may be obtained by increasing the ram pressure of the secondary relative to 
the primary. However if it is increased too much, the subcluster will actually
penetrate the core of the primary and the bar-like structure will be destroyed.
Similarly, decreasing the impact parameter may also increase the offset.

Since the merger is slightly offaxis, a portion of the subcluster misses
the primary core and meets with significantly less resistance as it
passes to the south of the X-ray peak. Figure \ref{simdat1}d shows the location of the 
respective gas components relative to the X-ray emission.   The red indicates
subcluster gas while black indicates primary cluster gas.
In this analysis,
we have used a dynamically inert quantity to trace the individual gas components
within the velocity field. We find that
the X-ray peak is  composed almost entirely of gas originating with the primary cluster. 
The bulk of the subcluster gas is located SE of the X-ray peak. Very
little mixing of the two components has occurred at this time. The leading
edge of the subcluster gas appears as the  westward
extension of red. As the subcluster gas  passes by the core,
it creates a shear which drags both primary and subcluster gas into
an extended tail along the bottom edge of the X-ray peak. The
overall morphology of the remnant X-ray core is seen to have an angled or
L-shaped distribution. This is reminiscent of a feature
identified in the wavelet analysis of A754's X-ray surface brightness
by  Slezak \etal (1994) (Figure 16, therein).  Along the interface
between the two gas components, we note a spray of gas both to the NE and SW,
essentially perpendicular to the merger axis. The velocity of
the spray is $\sim$1000 \kms near the X-ray peak but is seen to broaden and decelerate
as it exits the core.

Comparison of the location of the subcluster gas component in  Figure \ref{simdat1}d
with the location of the dark matter distribution in  Figure \ref{simdat1}b
reveals that the subcluster gravitational potential (NW) has been completely stripped
of its original gas content. A similar stripping was noted by Pearce \etal (1994). 
This is not to say that the remnant subcluster is
devoid of gas. Rather, it drags along gas originally associated with the primary causing
it to interact strongly with gas infalling from the outer regions of the cluster.
The interface  between the expanding bow shock and infalling gas is
most noticeable in the temperature distribution (Figure \ref{simdat1}a, \S \ref{tempdist})
and in the gas velocity vectors (Figure \ref{simdat1}c) where a sharp transition is seen
in the NW between gas moving to the NW at up to 1000 \kms and gas moving to the SE at
nearly 700 \kmsp  Regarding the X-ray morphology, it is the gas being dragged outward by 
the subcluster potential that accounts for the
overall (SE to NW) elongation of the X-ray surface brightness. The implications for
gas temperature morphology are discussed below (\S \ref{tempdist}).

Of course the gasdynamics in Figure \ref{simdat1}c would not be directly
observable since they are in the plane of the sky. In Figure \ref{vzscan},
we show the line-of-sight  gas velocities (vertical axis) in an east-west 
scan (horizontal axis) which intersects the X-ray surface brightness peak.
That is, Figure \ref{vzscan} represents gas velocities in a plane perpendicular
to the plane of the sky. Evident in this figure is a general expansion
of the cluster gas. The largest velocities ($\sim$1250 \kms) are in the region 
of the radial spray described above. We also see expansion velocities near 500 \kms in
the region of the bow shock (right hand side). These velocities should
be resolved by ASTRO-E provided the emissivity is great enough in the
high velocity regions. Of course in an actual spectroscopic observation,  the dynamical
structure seen in Figure \ref{vzscan} would be projected along the line-of-sight
and emission-weighted.

\subsection{Temperature Distribution}
\label{tempdist}

Figure \ref{simdat1}a shows the projected, emission-weighted temperature distribution
(color)
upon which is superimposed the X-ray surface brightness distribution (contours). The basic
temperature morphology consists of two hot spots. The smaller, lower
temperature hot spot located somewhat south and east of the
X-ray peak is formed when residual gas from the subcluster
 collides with gas stalled near the X-ray peak. As such, this feature
is not obviously apparent in the observational data although it
would be more consistent with the data if it were a somewhat lower temperature and/or further S/SW.
Our parameter space survey shows it to be a strong function of both the relative
gas content of the clusters and the impact parameter.   Decreasing
the gas content of the subcluster relative to the primary decreases
the hot spot temperature relative to the surrounding gas by decreasing
the ram pressure and limiting the residual infall. Decreasing the impact parameter moves the hot spot 
to the north and increases its temperature relative to the surrounding gas.
Therefore, a better fit to the data may be possible by slightly increasing
the impact parameter and decreasing the gas content of the subcluster.
However, this must be balanced by other features. Specifically,
we need sufficient gas in the subcluster to create the ram-pressure flattening
and displacement of the X-ray
peak (\S \ref{xmorph}). This would argue for a more concentrated gas distribution
in the subcluster.

The second hot spot appears as a large slightly
asymmetric arc near the western edge of the X-ray emission, Figure \ref{simdat1}a. 
The peak projected, emission-weighted temperature
is greater than 19 keV and largely coincides with the subcluster dark 
matter distribution. As the subcluster impinges on the primary a bow shock
forms which produces a conical sheet of extremely hot gas. As the merger
progresses, the shock heated gas is driven through the primary. When viewed
in projection, it appears as the arc that we see here.  Although
the subcluster is stripped of its initial gas content, the 
dark matter potential remains intact and drags a significant amount of
primary cluster gas with it. This gas continues to interact with gas in the outer
regions of the primary and helps to maintain the temperature of the
original shock heated gas. Coincident with the arc of hot gas is
a discontinuity in the gas velocities which approaches 1700 \kmsp

Another interesting temperature feature is the band of cooler gas ($\sim$7.5 keV) that 
runs along the X-ray peak.
This is a combination of pre-shocked primary cluster gas (initially 6.7 keV)
and subcluster gas that is deflected by the core and forced outward in  
the radial spray mentioned above (see \S \ref{xmorph}). As the spray
expands and decelerates, the temperature drops to $\sim$6.5 keV at
0.5 Mpc to the NE.

Figure \ref{tempcomp1} shows a quantitative point-by-point comparison
of our model with the ASCA-based temperature map of A754 (HM96). 
Since we find a general consistency between the ASCA-based and ROSAT-based (HB95)
temperature maps, we will only compare directly with the former. For the purposes of 
this analysis, we have
placed a grid similar in scale and location (with respect to the X-ray
emission) to the one used by HM96. Within each numbered region, we
have calculated the mean projected, emission-weighted temperature
in the model. We then compare these to the values observed in the
corresponding regions. Error bars on the observed values are
90\% confidence levels. Error bars on the model represent 1-$\sigma$
about the mean. Data points have been slightly offset to
minimize confusion. In each instance, the model is statistically
consistent with the observed temperatures. The largest absolute
discrepancy is found in region 7. We do find gas at
$\sim$19.5 keV in this region, however because of the exact location 
of the grid, it becomes somewhat diluted with adjacent, cooler gas.

In these data, we recognize three basic features. 1) The negative
gradient from region 1 to region 3. 2) The relatively isothermal
regions 4,5, and 6. 3) The negative gradient from region 7 to region 9.
Each of these features is well-represented in the model. The largest
discrepancy is found in regions 4-6 where the model shows a slightly
positive gradient and the data shows a somewhat stronger negative gradient
because of the relatively hot region 4.
Still, they are statistically consistent. As was suggested above, better
agreement in region 4 might be obtained if the hot spot near the
X-ray peak (Figure \ref{simdat1}c) were further S-SW.

Of course, the X-ray telescope only allows us to see the projected,
emission-weighted temperature. In doing so, much of the detailed
structure of and extremes in temperature can be hidden. Figure \ref{tempact}
shows the zone-by-zone distribution of the true (\ie non-emission weighted) 
model gas temperatures within
a cube comparable in dimension to that delineated by the 9 regions in Figure \ref{tempcomp1}.
Note the wide range in temperatures (5-22 keV). Although the bulk of
the volume is dominated by gas less than 10 keV, the distribution is
strongly skewed toward higher temperatures. Fully, 29\% of the volume
is in gas greater than 11 keV. The median temperature is 9.9 keV, the
mean temperature is 10.9 keV. Compare these values to the mean emission-weighted
temperature of 9 keV and the initial primary cluster temperature of
6.7 keV.

\section{FUTURE EVOLUTION}
\label{evol}
Figures \ref{simdat3} and \ref{simdat6} depict the model evolution at
2.75 and 5.25 Gyr after closest approach. These are the same quantities
represented in Figure \ref{simdat1} (T=0.3 Gyr) and can therefore be compared
directly. Considerable evolution is evident.
The subcluster dark matter distribution has proven to be
rather resilient. Not only does it survive the initial passage, but
a couple of others as well. Through examination of the particle group velocities,
we identify subsequent merger events at 1.6 and 2.5 Gyr. Although far less
dramatic than the initial merger, these events are still significant
in that they continually stir the ICM preventing a rapid relaxation.
At 2.75 Gyrs, the system is experiencing a maximum separation which results
in the peanut-shaped morphology of the dark matter distribution seen in
Figure \ref{simdat3}b (color). It should be noted that survival of substructure
in the dark matter will depend on the relative masses of the subclusters as
well as their relative concentration. As the mass ratio decreases (approaching 1:1),
the subcluster will survive longer. Also, chances of survival will be enhanced
as the subcluster mass distribution
becomes more concentrated and as the impact parameter increases.

The continued dynamical evolution
is still apparent in the X-ray surface brightness morphology (all 4 panels, Figure
\ref{simdat3}). Although considerably more symmetric (\ie relaxed) than during the
epoch we find most consistent with A754, there is still a significant extension to the SE which is abruptly
flattened. Comparison with the velocity vectors in Figure \ref{simdat3}c,
indicates that the extension results from a southerly flow  of $\sim$650 \kmsp
The flattening occurs in a region of strong shear where the southerly flow
is abruptly cutoff by a WSW flow of $\sim$700 \kmsp In fact, the WSW flow is
part of a more global circulation that is seen to extend unbroken (at least
in the plane of the merger) around the entire cluster. In this sense, the
region of shear described above is the location where the flow meets itself.
One by-product of the global circulation is the mixing of the subcluster
and primary cluster gas components. Figure \ref{simdat3}d demonstrates how the
global circulation has drawn the subcluster gas (red) around the
remnant core. We examine the rate of gas mixing more quantitatively in
\S \ref{mixing}  Finally, the temperature extremes seen in A754 have
been reduced significantly by 2.75 Gyrs (Figure \ref{simdat3}a). At this time, 
the cluster would be considered largely isothermal, at least within the central region. 
To the NW, we find large regions of relatively cool gas ($<$6 keV) which
are likely  remnants from the initial merger.  Although 
 radiative cooling was not included in these simulations, we do find
regions where the gas cools via adiabatic expansion such as behind the
initial bow shock.

Between 2.75 and 5.25 Gyr, there continues to be minor ``sloshing"
within the gravitational potential. The relative dark matter
center of mass velocities peak at only 100 \kmsp At 5.25 Gyr,
the X-ray surface brightness morphology is quite regular with
a slight elongation from  SE to the NW (Figure \ref{simdat6}).
 A similarly relaxed dark matter distribution is  apparent in
Figure \ref{simdat6}b (color). Here, the dark
matter distribution appears somewhat more elongated than
the gas distribution. Like the gas distribution, the 
dark matter major axis is oriented
SE to NW reflecting the axis of the secondary mergers. Unlike
the gas distribution, the dark matter can shed angular
momentum through free streaming particles. Consequently, the
secondary mergers discussed above are largely head-on.

Although temperature inhomogeneities persist, the overall
temperature distribution is noticeably more regular (Figure \ref{simdat6}) than
at earlier epochs.
The cool regions noted at 2.75 Gyr have vanished while
a  radial temperature gradient peaking at 11 keV in the core has developed. 
Pearce \etal (1994) also note a centrally peaked temperature distribution.
The relative heating in the core results from the continual stirring
of the relatively high density gas in the core by fluctuations in the gravitational
potential minimum as particles continue to stream back and forth along
the merger axis. In the absence of radiative cooling, there is no way to dissipate 
the heat. We also find that the region NW of the cluster core is 
somewhat warmer ($\sim$1-2 keV) than the corresponding region to the SE. 
The overall temperature morphology at this time is similar to that
of Triangulum Australis (Markevitch, Sarazin \& Irwin 1996).

Largescale gasdynamics (both bulk flows and rotation) persist
even as late as 5.25 Gyr.  Still evident by the velocity vectors
in Figure \ref{simdat6}c is a general counterclockwise circulation
in the merger plane.  Comparing to Figure \ref{simdat3}c, we
note that the flow is less coherent and of significantly reduced
magnitude. The highest velocity gas ($\sim$850 \kms) is located in the extreme
SE corner of Figure \ref{simdat6}c. Within the central cluster,
gas velocities range up to 600 \kmsc but the average in nearer
300 \kmsp  As mentioned above, the gas is less
efficient than the dark matter  at shedding angular momentum.
Over time, angular momentum is transferred via mixing to adjacent
gas components. Similarly, kinetic energy is 
transferred from one component to the next and ultimately converted
into heat. The relative inefficiency of this process is similarly apparent
in the slow rate of mixing between primary and subcluster
gas components, particularly in the outer regions of the merger remnant (Figure 
\ref{simdat6}d). We address mixing in more detail below, \S\ref{mixing}.

\subsection{Evolution of Dynamical Pressure Support}
\label{pressure}

Since dynamical information regarding the X-ray emitting
gas is non-existent at this time, the analysis of 
clusters has always assumed them
to be thermally supported (\ie in hydrostatic equilibrium). Several
recent cluster analyses have argued for  additional pressure components (\eg
dynamical, magnetic fields)
in order to explain the apparent discrepancies between X-ray, virial and gravitational
lensing based cluster mass estimates (\eg Loeb \& Mao 1994, among others). Using our model of A754,
we can determine the relative  significance  of dynamical pressure
support throughout the merger evolution.

  Figure \ref{pressrat} shows the evolution
of $P_K$/$P_T$ ($P_K=\rho v^2$  and $P_T=kT$) within spherical volumes (r=0.4 and 1.5 Mpc)
centered on the gravitational potential minimum.  Initially, the dynamical
pressure is negligible although it grows steadly as the subcluster impinges
on the primary.  A maximum is reached shortly before closest approach (T=0) when
the dynamical pressure is greater than 60\% of the thermal pressure 
within 0.4 Mpc.  Secondary maxima occurring near 2 and 3 Gyr correspond
to subsequent passages by the subcluster dark matter remnant.  As might be 
expected, the larger volume (r=1.5 Mpc) experiences fewer extremes than the core
which may explain the relative heating in the core. After the initial impact,
the dynamical pressure is typically $\sim$20\% of the thermal pressure but
decays to 5-10\% after 4.5 Gyrs. Therefore, late in the merger evolution,
gasdynamics play a limited role in supporting the expanded gas core. This
is consistent with Pearce \etal (1994).

Figure \ref{pressrat} has implications for X-ray based cluster mass estimates. 
 Neglecting the dynamical component of the pressure support will cause
an underestimation of the cluster mass by a fraction  
1-(1+($P_K$/$P_T$))$^{-1}$.
Based on this analysis, we would predict errors of $\sim$20-40\% or more during the
2 Gyrs immediately after a merger and errors of 10-20\% at later times.
Note also the radial dependence. Mass estimates based on only the central
regions of the cluster will be more severely affected. These results are consistent, at least
in magnitude, with previous studies (\eg Evrard, Metzler \& Navarro 1996; 
Schindler 1996; Roettiger \etal 1996)
although none of these studies showed evidence for systematic underestimation of the cluster masses
indicating that other effects (\eg projection effects) and the details of the 
observational analysis are also important.
By underestimating the total mass, the baryon fraction would be overestimated by the same factor.
Although significant, the errors noted here are not enough to account for
the apparent inconsistency between measurements of the baryon fraction in
clusters and that derived from primordial nucleosynthesis models that assume $\Omega$=1
 (White \etal 1993). Of course lowering the assumed value of $\Omega$ also reduces
the perceived discrepancy.

\subsection{Angular Momentum Evolution}
\label{bigmo}

The offaxis merger imparts considerable and sustained  angular momentum upon
the gas distribution in the merger remnant (Fig \ref{simdat6}). 
Although
the gas distribution is not rotationally supported, it is
slightly less prolate than the
dark matter distribution where angular momentum is dissipated
more readily by free-streaming particles.  Figure \ref{angmo} shows the evolution
of total angular momentum density ($L_{tot}$) within concentric spheres of
radius 0.4 and 1.5 Mpc centered on the gravitational potential 
minimum. We define $L_{tot}=\left [ \Sigma_i^3 (L_i)^2 \right ]^{0.5}/N_z$
and $L_i=\Sigma_j^{N_z} ({\bf r}_j \times \rho_j {\bf v}_j)_i$ where
${\bf r}_j$ is the distance from the potential minimum to the $j$th zone,
$\rho_j$ and ${\bf v}_j$ are the gas density and velocity, respectively,
 within the $j$th zone, and $N_z$ is the number of zones within the volume.
The sum over $i$ represents the sum over the Cartesian coordinates.

The angular momentum evolution in Figure \ref{angmo} closely mimics the evolution of the
non-thermal pressure support (see \S \ref{pressure}, Figure \ref{pressrat}).
The abrupt rise at T=0 (R $<$ 0.4 Mpc) occurs as the subcluster passes
near the primary core. The subsequent decay occurs as the subcluster leaves
the sphere only to return at 1.5-2.0 Gyr. Beginning at $\sim$3.5 Gyr,
the angular momentum within 0.4 Mpc is seen to decay steadily as the
core relaxes and angular momentum is transferred via mixing to larger 
radii. Within the larger volume, the total angular momentum
remains fairly constant, peaking slightly at 1.5-2.0 Gyr. Once
again, this shows that although the canonical core sound crossing time 
(and, therefore its relaxation time) is of order 1 Gyr, true relaxation
of the merger remnant does not really begin until after $\sim$3 Gyr.
The longer relaxation time is important to the interpretation of
substructure frequency. A longer relaxation time allows for greater consistency
between a high substructure frequency and a low $\Omega$, the density of the universe
relative to the critical density (Boute \& Xu 1997).

\subsection{Mixing}
\label{mixing}

Using a dynamically inert quantity evolved within the velocity field,
we have traced the rate at which the primary and subcluster gas components
mix. In addition to the non-isothermal temperature distributions observed
in some clusters, Arnaud \etal (1994) have also noted an inhomogeneous abundance
distribution in Perseus.  Since considerable scatter is observed in global
cluster abundances (Mushotzky \& Loewenstein 1997), it may be reasonable to expect inhomogeneous distributions
within individual clusters to arise as a result of mergers between
clusters of differing initial abundances. It then becomes important to quantitatively
assess the rate at which the remnant gas distribution becomes homogeneous.
A cursory examination of Figures \ref{simdat1}d, \ref{simdat3}d, and \ref{simdat6}d
indicates that mixing is slow in any case but even more so at larger radii
where the gasdynamics are less extreme.  Even after 5.25 Gyrs, there is
a noticeable patchiness in the gas distribution. This result may be sensitive
to cosmological infall which we do not simulate.

Figure \ref{color} presents the degree of mixing within spheres of radius
0.4 and 1.5 Mpc. Plotted is the fraction of zones within the corresponding
volumes that are at least 10\% mixed. Mixing does not begin until
after the time of closest approach (T=0) at which point the degree
of mixing within both 0.4 and 1.5 Mpc increases steadily  with time.
The level of mixing is consistently lower in the larger volume although
the rate of mixing within the two volumes track well until about 3-4 Gyrs
after the initial passage.
At this time, mixing within the 1.5 Mpc sphere effectively stops
while the rate of mixing within 0.4 Mpc actually increases. Recall from
\S\ref{bigmo} that it is at 3.5 Gyr when the ordered circulation in
the core gives way to random motions thus increasing the mixing efficiency.
After 5 Gyrs, 90\% of the zones within 0.4 Mpc are at least 10\% mixed which
compares to only 45\% within the larger volume. This analysis would seem to
indicate 1) a tendency to find more homogeneous gas distributions within the cores
of clusters, 2) the potential for abundance gradients within
clusters having evolved sufficiently since their last significant merger, and 
3) the potential for patchiness in the distribution of metals in the
outer regions of clusters.  Of course a patchiness may also  result from
a local enhancement of metals by supernovae in individual galaxies.

\section{SUMMARY}
\label{summary}

We have presented a dynamical 3-dimensional model of A754 which can explain
many of its observed  morphological peculiarities. In our model, A754
is the result of a very recent ($<$0.3 Gyr), nearly in the plane-of-the-sky merger
 between two clusters having a total mass ratio that is likely less than 2.5:1.
The merger is slightly off-axis, with an impact parameter of 120 kpc or about
half the initial primary cluster core radius. The final impact velocity
is $\sim$2500 \kmsp  Although the model is poorly constrained by the 
limited observational data, our assumed initial conditions for the merger are consistent with
that data. The fact that the observed characteristics of A754 can be reproduced so well
suggests the actual properties of the merging clusters in A754 cannot be too
different from those adopted here. Moreover, we believe that the dynamics described here 
are largely representative of slightly off-axis mergers. 

In general, the results of these simulations are consistent with other studies
of major cluster mergers ( Roettiger \etal 1993, 1996, 1997a; Schindler \& M\"uller 1993; Pearce \etal 1994).
In particular, they reveal how gravitational energy is transferred to the ICM
and dissipated via shocks, resulting in a heating
and expansion of the ICM relative to that of the dark matter, and  high velocity bulk 
flows within the gas. It should be noted
however that in this study, as well as many others, potentially important physical 
processes (\eg radiative cooling, thermal conduction, magnetic fields, etc.) have been ignored.
Moreover, we study the merger of isolated subclusters in hydrostatic equilibrium rather than
mergers within  an evolving largescale structure. On the other hand, this allows us to focus
our numerical resources on the dynamics of the merger itself.  In addition, we are able to systematically
identify and analyze the physical processes which govern the dynamics during mergers.
Thus, our results have important implications for the dynamics of off-axis mergers in
general. 

More specifically, the main conclusions of this study include:

 1) Major merger events can account
for many if not all of the morphological irregularities observed in A754
as well as other well-resolved clusters. The extreme temperature inhomogeneities
of the type observed in A754 are currently the most direct indicator of
recent dynamical evolution and are particularly useful for ruling out more
subtle projection effects. The simple projection of two clusters along a common
line-of-sight cannot account for the sharp temperature transitions, due to shocks,
nor the extreme nature of the observed temperatures (19 keV).

2) Relaxation of the merger remnant is delayed beyond
the canonical sound crossing time by subsequent passages of the subcluster's
remnant dark matter. The extended relaxation time allows for greater consistency between a high
frequency of substructure in clusters of galaxies and a low $\Omega$ universe (Buote \& Xu 1997).

3) Major merger events can result
in significant non-thermal pressure support within the remnant gas distribution.
The existence of a non-thermal pressure component
may seriously affect X-ray based cluster mass estimates which often
employ the assumption of hydrostatic equilibrium. Errors are lower when the mass estimates
are made at larger radii, but they can still approach
30\% within 1.5 Mpc during the early stages of a merger. At later times, errors are more typically 15\%. 
These results are consistent in magnitude with Evrard \etal (1996) and Schindler (1996).

4) Off-axis mergers can impart significant angular momentum on to
the remnant gas distribution resulting in a sustained rotation about 
the cluster core. The magnitude of rotation (and bulk flows) seen in this model should be
observable with ASTRO-E. Although much of the gasdynamics in  A754 merger are not directly
observable since they occur in the plane of the sky, we do predict that
a general expansion of the cluster ICM may be  observable provided the emissivity of the high
velocity gas is great enough, see \S\ref{xmorph}  Since the time period between major
mergers is likely comparable to or less than the observed relaxation time scales, the
gasdynamics within the cluster may largely reflect those imparted by the last major merger.

5) Mixing of premerger gas components is an inefficient
process and may explain the observed patchiness and gradients in both temperature and
abundance distributions (\eg Perseus, Arnaud \etal 1994; Coma, Honda \etal 1996).
Mixing is most efficient in the core of the cluster where the gasdynamics are most significant.
\bigskip

We thank Jack Burns for many useful discussions. We also thank the anonymous
referee for a careful reading of the manuscript and many useful suggestions.
Computations were performed on the MasPar-2 at the Earth and Space Data
Computing Division of Goddard Space Flight Center and on the CM-5 at
the National Center for Supercomputing Applications. KR thanks the
National Academy of Sciences/National Research Council for financial
support. This research has made use of the NASA/IPAC extragalactic database
(NED) which is operated by the Jet Propulsion Laboratory, Caltech, under contract
with the National Aeronautics and Space Administration.

\newpage

\onecolumn
\begin{table}
\label{tab1}
\begin{center}
\begin{tabular}{c c c c c c c  } 
\multicolumn{7}{c}{Table 1.  Observed Subcluster Galaxy Parameters}\\ \hline \hline
\multicolumn{1}{c}{ }&
\multicolumn{2}{c}{$v_r$ (\kms)} &
 \multicolumn{2}{c}{$\sigma_v$ (\kms)} &
 \multicolumn{2}{c}{$N_g^1$} \\ \hline
\multicolumn{1}{c}{ }&
 \multicolumn{1}{c}{SE$^2$} &
 \multicolumn{1}{c}{NW$^3$} &
 \multicolumn{1}{c}{SE} &
 \multicolumn{1}{c}{NW} &
 \multicolumn{1}{c}{SE} &
 \multicolumn{1}{c}{NW}\\ \hline
F86$^4$ & 16145$\pm$160  & 16165$\pm$217  & 733$\pm^{156}_{102}$ & 839$\pm^{223}_{132}$ & 21    &  15 \\
EM92$^5$ & 16387    & 16007    & 698  & 521  & 12  & 10  \\
ZZ95$^6$ & 16320$\pm132$  & 16432$\pm191$  &   830$\pm^{130}_{112}$ & 970$\pm^{156}_{134}$ &  41  & 27  \\ \hline 
\end{tabular}
\end{center}

\caption[Sample Parameters]
{$^1$ Number of galaxies associated with the subcluster
$^2$ Southeastern subcluster,
$^3$ Northwestern subcluster,
$^4$ Fabricant \etal 1986,
$^5$ Escalera \& Mazure 1992,
$^6$ Zabludoff \& Zaritsky 1995}

\end{table}

\begin{table}
\label{tab2}
\begin{center}
\begin{tabular}{c c c c c c c c c c} 
\multicolumn{10}{c}{Table 2. Initial Cluster Parameters}\\ \hline \hline

 \multicolumn{1}{c}{Cluster$^1$} & 
 \multicolumn{1}{c}{$M_{tot}^2$} &
 \multicolumn{1}{c}{$T_e^3$} &
 \multicolumn{1}{c}{$\sigma_v^4$} &
 \multicolumn{1}{c}{$r_c^5$} &
 \multicolumn{1}{c}{$f_g^6$} &
 \multicolumn{1}{c}{$\beta^7$}&
 \multicolumn{1}{c}{$n_{eo}^8$} &
 \multicolumn{1}{c}{$v_{impact}^9$} &
 \multicolumn{1}{c}{$\Delta r^{10}$} \\

 \multicolumn{1}{c}{ ID } & 
 \multicolumn{1}{c}{ (10$^{14}$ M$_\odot$) } &
 \multicolumn{1}{c}{ (keV) } &
 \multicolumn{1}{c}{ (km s$^{-1}$) } &
 \multicolumn{1}{c}{ (kpc) } &
 \multicolumn{1}{c}{     } &
 \multicolumn{1}{c}{     } &
 \multicolumn{1}{c}{ (10$^{-3}$cm$^{-3}$) } &
 \multicolumn{1}{c}{ (km s$^{-1}$)} &
 \multicolumn{1}{c}{ (kpc)}  \\ \hline

1 & 8.0 & 6.7 & 785  & 220 & 0.12 & 0.75 & 1.55 & 2500 & 120 \\
2 & 3.2 & 3.1 & 526  & 135 & 0.06 & 0.72 & 1.01 & & \\ \hline 
\end{tabular}
\end{center}
{ $^1$ Cluster ID. 
 $^2$ Total Mass R$<$3 Mpc.
 $^3$ Temperature.
 $^4$ Velocity Dispersion. \\
 $^5$ Core Radius. 
 $^6$ Global Gas Fraction, by mass. 
 $^7$ $\beta$-parameter.\\
 $^8$ Central Gas Density. 
 $^9$ Impact Velocity (dark matter).
 $^{10}$ Impact Parameter.}
\end{table}

\begin{table}
\label{tab3}
\begin{center}
\begin{tabular}{c c c c c c c} 
\multicolumn{7}{c}{Table 3. Post-Merger Comparison}\\ \hline \hline

 \multicolumn{1}{c}{  } & 
 \multicolumn{1}{c}{N$_g$(SE)$^1$} &
 \multicolumn{1}{c}{N$_g$(NW)$^2$} &
 \multicolumn{1}{c}{$\sigma_v$(SE)$^3$} &
 \multicolumn{1}{c}{$\sigma_v$(NW)$^4$} &
 \multicolumn{1}{c}{$<T_e>^5$} &
 \multicolumn{1}{c}{$M_{tot}^6$} \\

 \multicolumn{1}{c}{  } & 
 \multicolumn{1}{c}{  } &
 \multicolumn{1}{c}{  } &
 \multicolumn{1}{c}{ (km s$^{-1}$) } &
 \multicolumn{1}{c}{ (km s$^{-1}$) } &
 \multicolumn{1}{c}{ (keV)    } &
 \multicolumn{1}{c}{ (10$^{14}$ M$_\odot$ )   } \\ \hline

Observed & 41 & 27  & 830$\pm^{130}_{112}$  &  970$\pm^{156}_{134}$ &  9.0  & 10.48$\pm3.27$ \\
Simulated & 50000  & 20000  & 1034$\pm{105}$  & 830$\pm{145}$ & 9.1  & 9.5   \\ \hline 
\end{tabular}
\end{center}

{$^1$ Galaxy/particle count in SE subcluster (ZZ95).\\
 $^2$ Galaxy/particle count in NW subcluster (ZZ95).\\
 $^3$ Velocity dispersion in SE (ZZ95).\\ 
 $^4$ Velocity dispersion in NW (ZZ95). \\
 $^5$ Mean temperature (HM96).\\
 $^6$ Total mass within $\sim$2 Mpc (Escalera \etal 1994, virial mass).}
\end{table}

\newpage

\section*{ REFERENCES}
 \everypar=
   {\hangafter=1 \hangindent=.5in}

{
Arnaud, K. A., Mushotzky, R. F., Ezawa, H., Fukazawa, Y., Ohashi. t.,
Bautz, M. W., Crewe, G. B., Gendreau, K. C., Yamashita, K., Kamata, Y.,
\& Akimoto, F. 1994, ApJ, 436, L67

Arnaud, K. A. 1996, ADASS, A.S.P. Conference Series, eds. Jacoby, G. \& Barnes J., 101, 17

Bahcall, N. \& Lubin, L. 1994, ApJ, 426, 513

Binney, J. \& Tremaine, S. 1987, Galactic Dynamics, (Princeton: Princeton University Press)

Bird, C. 1994, AJ, 107, 1637

Biviano, A., Girardi, M, Giuricin, G., Mardirossian, F., \& Mezzetti, M. 1993, ApJ, 411, L13

Boute, D. A. \& Xu, G. 1997, MNRAS, 284, 439

Cen, R. 1992, ApJS, 78, 341

Colella, P. \& Woodward, P., 1984, J. Comp. Phys., 54, 174

Crone, M. M., Evrard, A. E., \& Richstone, D. O. 1994, ApJ, 434, 402

Davis, D. 1994, Ph. D. Thesis, University of Maryland, College Park

Dressler, A. 1980, ApJS, 42, 565

Dressler A. \& Shectman, S. 1988, AJ, 95, 985

Edge, A. C., Stewart, G. C., \& Fabian, A. C. 1992, MNRAS, 258, 177

Escalera, E. \& Mazure, A. 1992, ApJ, 388, 23 (EM92)

Escalera, E., Biviano, A., Girardi, M., Giuricin, G., Mardirossian, F., Mazure, A., \& Mezzetti, M. 1994, ApJ, 423, 539.

Evrard, A. E. 1990, ApJ, 363, 349

Evrard, A. E., Metzler, C. A., \& Navarro, J. F. 1996, ApJ, 469, 494

Fabricant, D., Beers, T. C., Geller, M. J., Gorenstein, P., Huchra, J. P., \& Kurtz, M. J. 1986 ApJ, 308, 530 (F86)

Geller, M. \& Beers, T. C. 1982, PASP, 94, 421

G\'omez, P., Loken, C., Burns, J. O. \& Roettiger, K. 1997, in preparation

Heckman, T., Baum, S., Van Breugel, W., \& McCarthy, P. 1989, ApJ, 338, 48

Henriksen, M. J. 1993, ApJ, 414, L5

Henriksen, M. J. \& Markevitch, M. L. 1996, ApJ, 466, L79 (HM96)

Henry, J. \& Briel, U. 1995, ApJ, 443, L9 (HB95)

Hockney R. W. \& Eastwood J. W. 1988, Computer Simulations Using Particles, 
(Philadelphia:IOP Publishing)

Honda, H., Hirayama, M., Wantanabe, M., Kunieda, H., Tawara, Y., Yamashita, K., Ohashi, T., Hughes, J. P., \&
Henry, J. P. 1996, ApJ, 473, L71

Jackson, J. D. 1975, Classical Electrodynamics, (New York:Wiley)

Jones, C. \& Forman, W. 1984, ApJ, 276, 38

Jones, C. \& Forman, W. 1991 BAAS, 23, 1338

Loeb, A. \& Mao, S. 1994, ApJ, 435, L109

Lubin, L. \& Bahcall, N. 1993, ApJ, 415, L17

Lucy, L. B. 1977, AJ, 82, 1013

Markevitch, M., Sarazin, C. L. \& Irwin, J. A. 1996, ApJ, 472, L17

Mazure, A., Katgert, P., Den Hartog, R., Biviano, A., Dubath, P., Escalera, E., Focardi, P.
Gerbal, D., Giuricin, G., Jones, B., Lefevre, O., Moles, M., Perea, J.
\& Rhee, G. 1996 ,A\&A, 311, 95

Mohr, J. J., Evrard, A., Fabricant, D. G., \& Geller, M. J. 1995, ApJ,
447, 8

Mushotzky, R. F. \& Loewenstein, M. 1997, ApJ Letters, in press

Navarro, J. F. \& White, S. D. M. 1994, MNRAS, 267, 401

Pearce, F. R., Thomas, P. A., \& Couchman, H. M. P. 1994, MNRAS, 268, 953

Ricker, P. M. 1997, preprint

Roettiger, K., Burns, J. O. \& Loken, C. 1993, ApJ, 407, L53

Roettiger, K., Burns, J. O., \& Pinkney, J. 1995, ApJ, 453, 634

Roettiger, K., Burns, J. O., \& Loken, C. 1996, ApJ, 473, 651

Roettiger, K., Loken, C., \& Burn, J. O. 1997a, ApJS, 109, 307

Roettiger, K., Stone, J. M., \& Mushotzky, R. F. 1997, ApJ, 482, 588

Schindler, S. 1996, A\&A, 305, 858

Schindler, S. \& M\"uller, E. 1993, A\&A, 272, 137

Slezak, E., Durret, F., \& Gerbal, D. 1994, AJ, 108, 1996

Stone, J. M. \& Norman, M. 1992, ApJ, 390, L17

West, M. J. \& Bothun, G. D. 1990 ApJ, 350, 36

White, D. A. \& Fabian, A. C. 1995, MNRAS, 273, 72

White, S.D.M., Navarro, J. F., Evrard, A. E., \& Frenk, C. S. 1993
Nature, 366, 429.

Zabludoff, A. I. \& Zaritsky, D. 1995, ApJ, 110, 447, L21 (ZZ95)

\newpage

\begin{center}{\bf FIGURE CAPTIONS}
\end{center}

{\bf Fig. \ref{obsdata}:} ROSAT X-ray surface brightness (solid
contours) and projected, galaxy density (dashed contours).
The galaxy distribution is derived from Fabricant \etal (1986) and
Dressler \& Shectman (1988). In all, 142 galaxy positions were smoothed
with 2.5\amin Gaussian. The X-ray surface brightness is smoothed
with a 40\asec Gaussian. Contours are at 0.08, 0.16, 0.24, 0.32, 0.4, 0.48,
0.56, 0.64, 0.72, 0.80, 0.90, 0.99 of the peak. 

{\bf Fig. \ref{simdat1}:} Simulated data at 0.3 Gyr after
closest approach. Each panel is 2 Mpc $\times$ 2 Mpc. The
smaller of the two clusters (subcluster) entered from the
left and is moving to the right.
a) X-ray surface brightness (contours)
and projected, emission-weighted temperature (color). 
Note the color bar above the panel.
b) Projected dark matter particle distribution (color)
c) Gas velocity within the plane of the merger. Vectors
are spaced at 75 kpc ( one third the resolution of the simulation)
and scaled to the maximum velocity indicated below the panel.
d) A dynamically inert quantity used to trace the subcluster
gas component (red) as it mixes with the primary
gas (black). Intermediate colors indicate the degree of mixing
as shown in the color bar below the panel. The X-ray contours
are at 0.03, 0.08, 0.16, 0.24, 0.32, 0.4, 0.48,
0.56, 0.64, 0.72, 0.80, 0.90, 0.99 of the peak. These are the
same as in Fig. \ref{obsdata} with the exception of the lowest
contour.

{\bf Fig. \ref{vzscan}:} Line-of-sight gas velocities in
an east-west scan that intersects the X-ray surface brightness peak.
The vertical axis represents the observer's line-of-sight looking into
the cluster. The horizontal axis
represents an east-west scan across the projected face of the cluster.  Contours are $\pm$250,$\pm$500,$\pm$750,
$\pm$1000,$\pm$1250 \kmsp Negative velocities (approaching)
are represented by dashed contours. Positive velocities (receding) are 
represented by solid contours. Tick marks represent 0.125 Mpc.
The plot dimensions are the same as the panels in Fig. \ref{simdat1}.
In a spectroscopic observation of the cluster, the velocity structure
seen in this figure would be emission-weighted and projected onto 
the plane of the sky. Consequently, much of the detailed structure may be 
unobservable.

{\bf Fig. \ref{tempcomp1}:} A comparison with the HM96
temperature map. a) Simulated X-ray surface brightness (contours)
and grid similar in scale and location (relative to the X-ray image)
as that used by HM96. b) Simulated projected, emission-weighted
temperatures ($\ast$) within the numbered regions in (a) compared
directly with the corresponding temperatures observed by HM96 ($\diamond$).

{\bf Fig. \ref{tempact}:} A histogram showing the zone-by-zone
distribution of temperatures within the volume delineated
by the nine regions in Figure \ref{tempcomp1}. 

{\bf Fig. \ref{simdat3}:} Same as Figure \ref{simdat1} only
at 2.75 Gyr after closest approach.

{\bf Fig. \ref{simdat6}:} Same as Figure \ref{simdat1} only
at 5.25 Gyr after closest approach.

{\bf Fig. \ref{pressrat}:} The evolution of the ratio of
dynamical to thermal pressure support within spheres of
radius 0.4 Mpc (solid line) and 1.5 Mpc (dashed line)
centered on the gravitational potential minimum. Times
are relative to the time of closest approach. This analysis
was performed on the low resolution simulation (\S \ref{num}).

{\bf Fig. \ref{angmo}:} The evolution of the angular momentum
density per zone relative to the gravitational potential
minimum within spheres of radius 0.4 Mpc (solid line) and
1.5 Mpc (dashed line) centered on the potential minimum. Times
are relative to the time of closest approach. This analysis
was performed on the low resolution simulation (\S \ref{num}).

{\bf Fig. \ref{color}:} The fraction of zones within spheres
of radius 0.4 Mpc (solid line) and 1.5 Mpc (dashed line)
centered on the gravitational potential minimum  which
exhibit at least 10\% mixing. That is, the gas within that
zone consists of at most 90\% of either primary or subcluster
gas. This analysis was performed on the low resolution simulation 
(\S \ref{num}).

\onecolumn
\begin{figure}[htbp]
\centering \leavevmode
\epsfxsize=0.9\textwidth \epsfbox{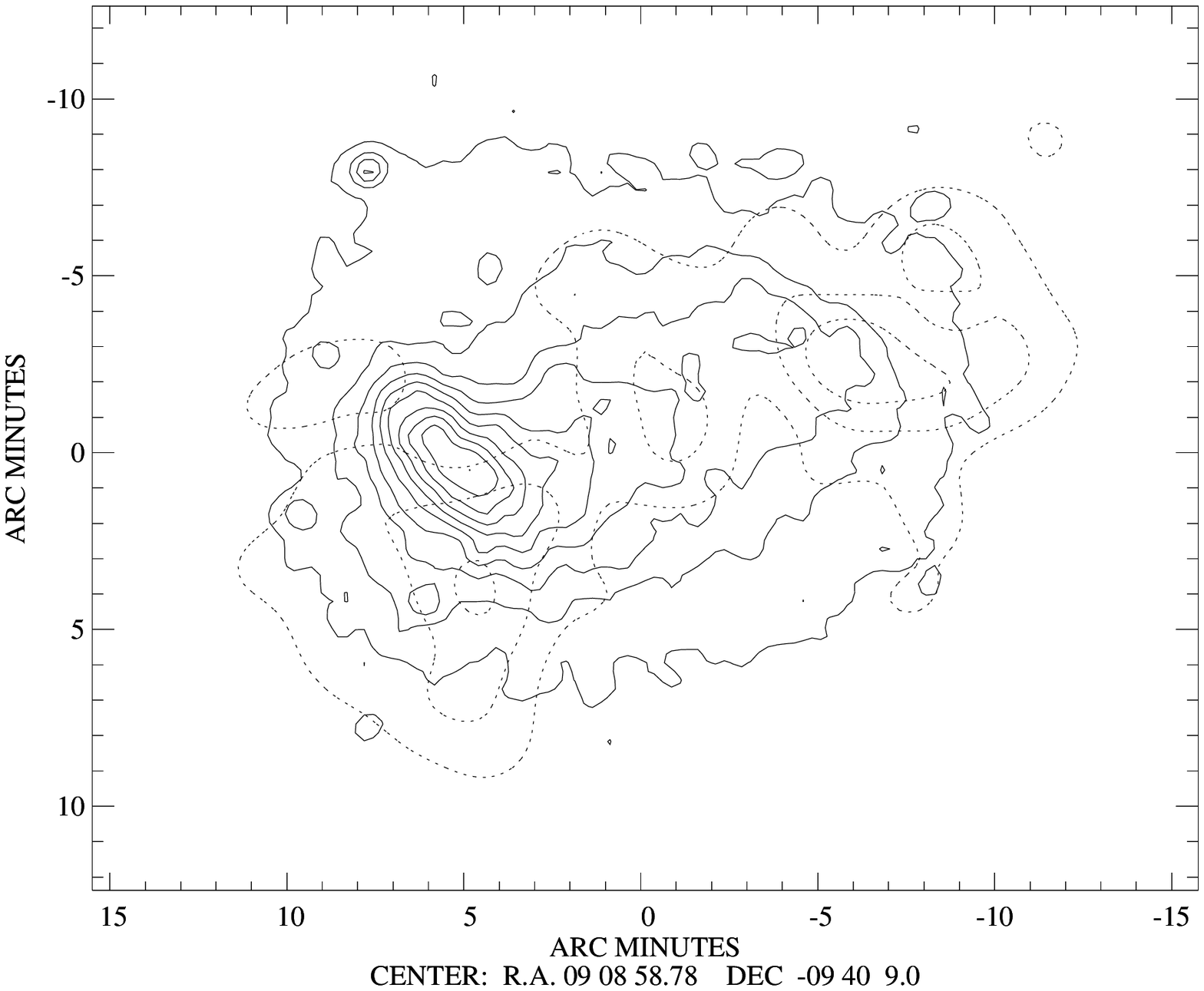}
\caption[]
{ }
\label{obsdata}
\end{figure}

\begin{figure}[htbp]
\centering \leavevmode
\epsfxsize=0.9\textwidth \epsfbox{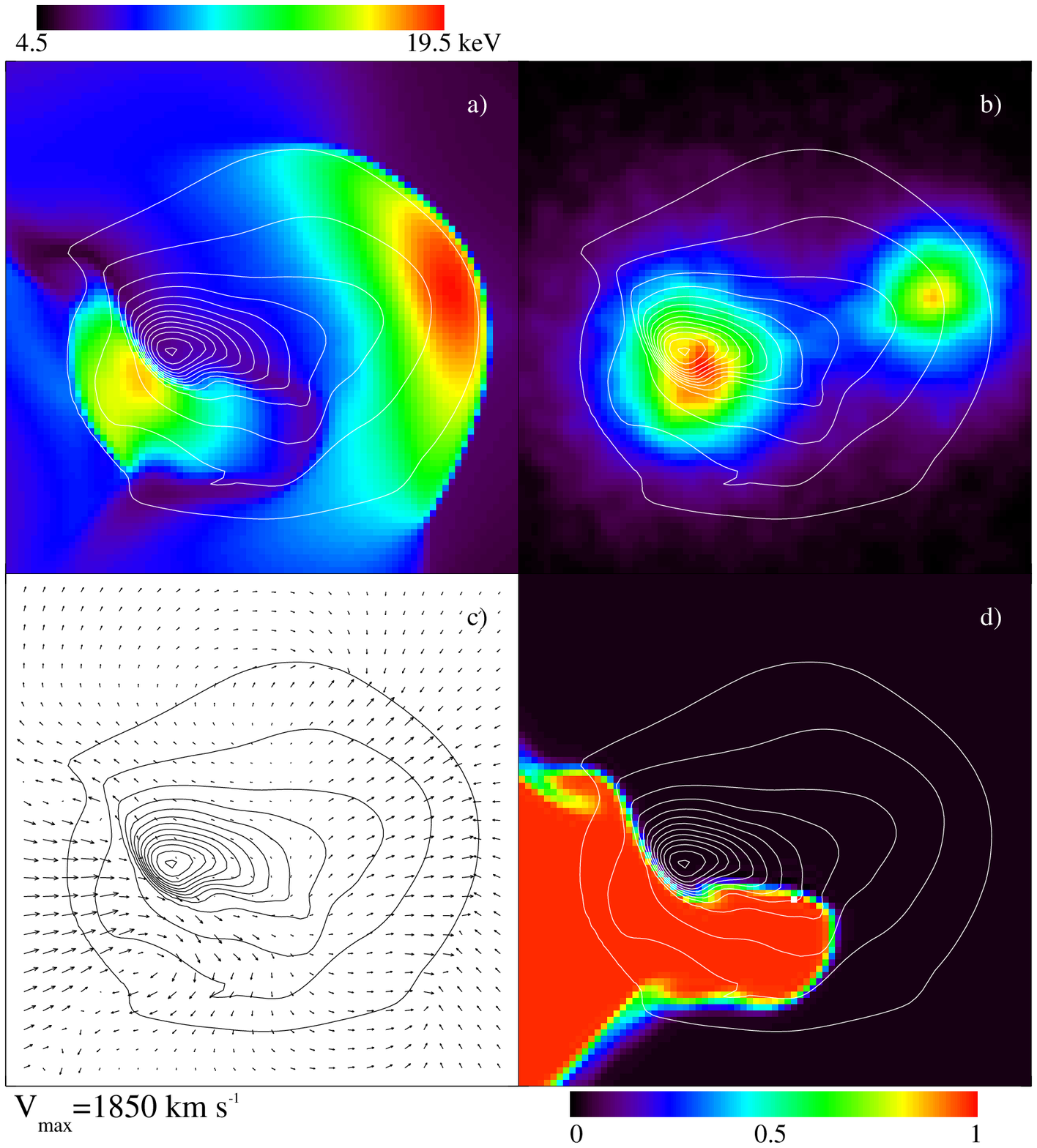}
\caption[]
{ }
\label{simdat1}
\end{figure}

\begin{figure}[htbp]
\centering \leavevmode
\epsfxsize=.5\textwidth \epsfbox{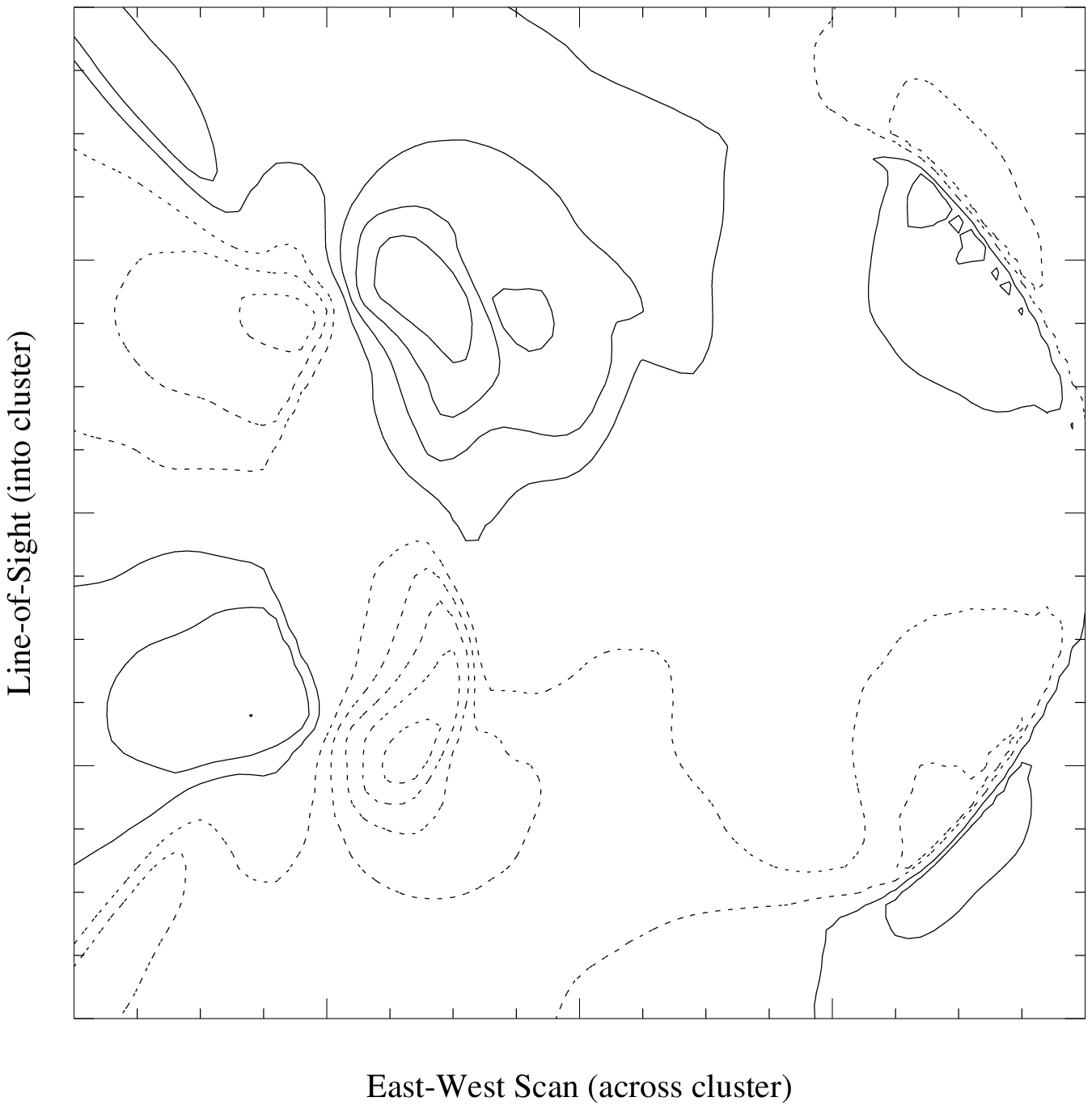}
\caption[]
{ }
\label{vzscan}
\end{figure}

\begin{figure}[htbp]
\centering \leavevmode
\epsfxsize=0.7\textwidth \epsfbox{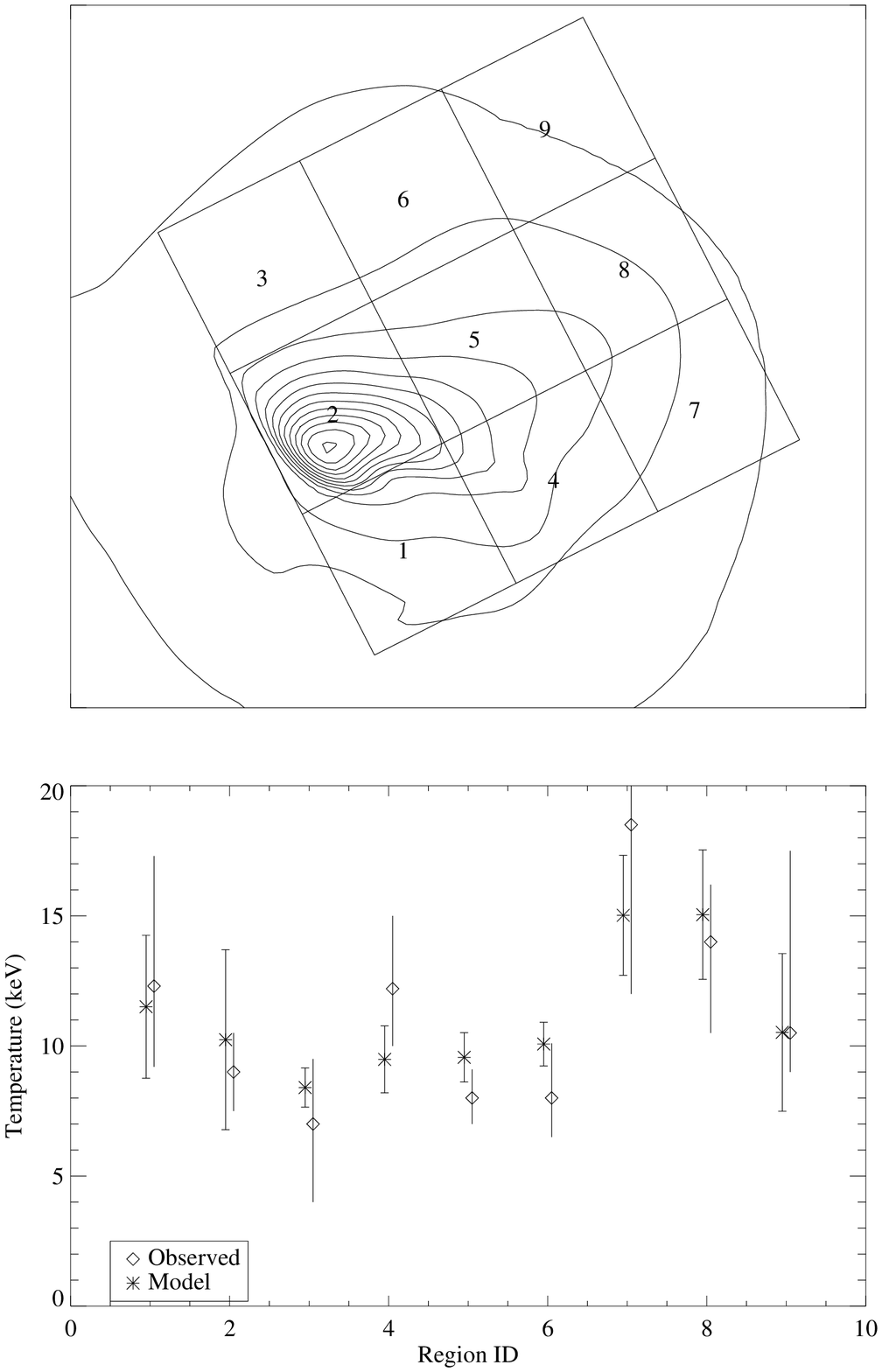}
\caption[]
{ }
\label{tempcomp1}
\end{figure}

\begin{figure}[htbp]
\centering \leavevmode
\epsfxsize=0.9\textwidth \epsfbox{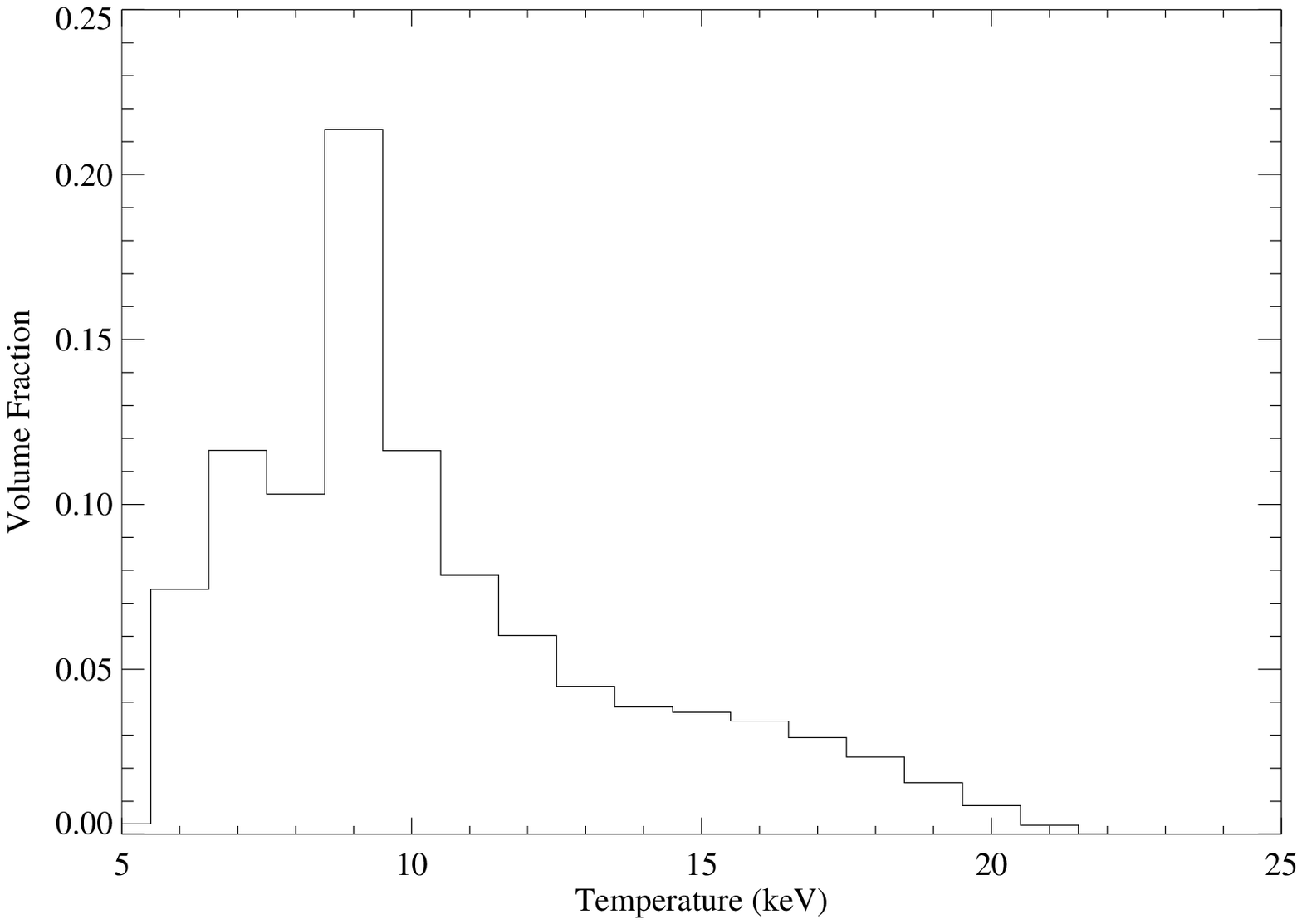}
\caption[]
{ }
\label{tempact}
\end{figure}

\begin{figure}[htbp]
\centering \leavevmode
\epsfxsize=0.9\textwidth \epsfbox{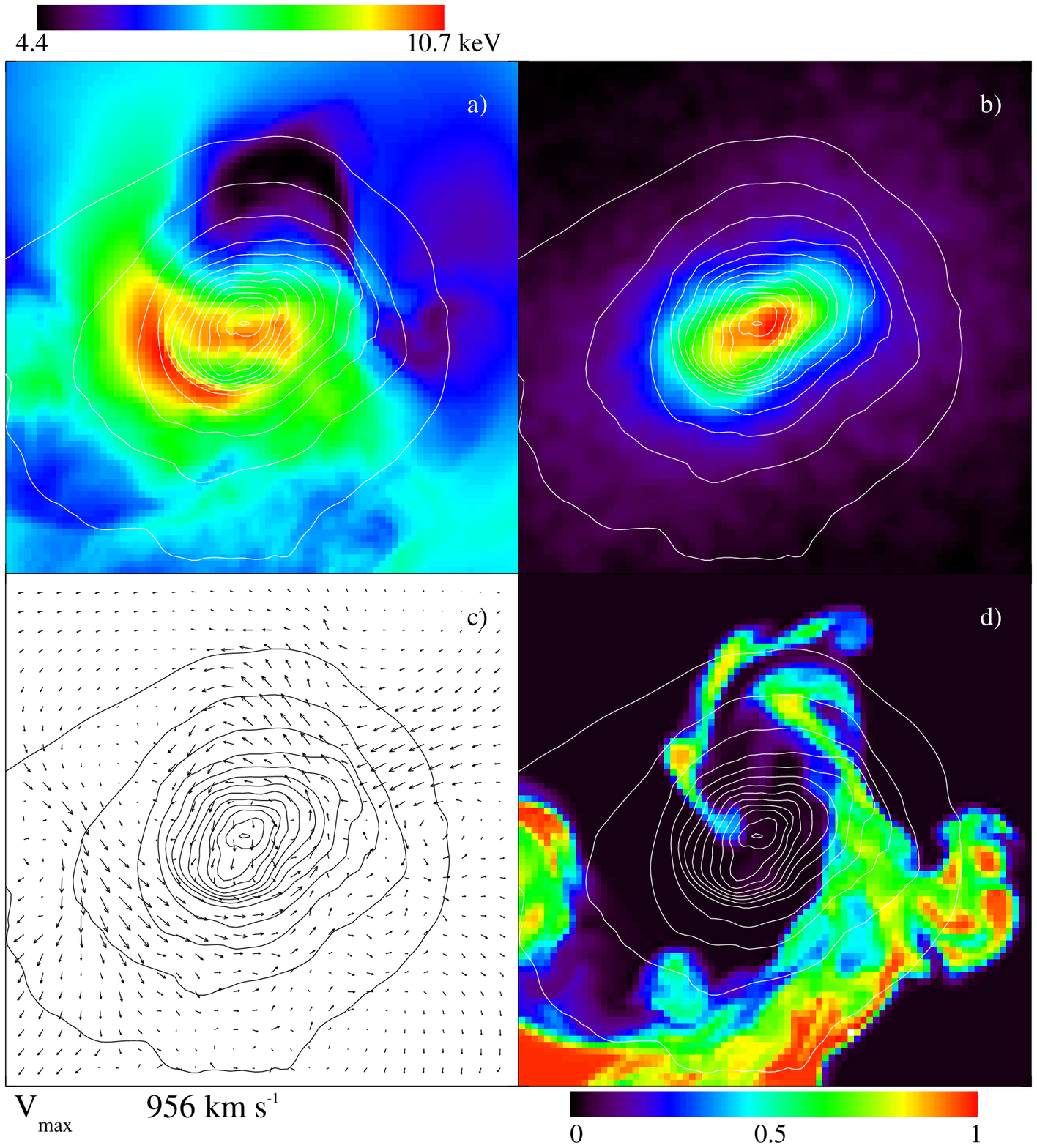}
\caption[]
{ }
\label{simdat3}
\end{figure}

\begin{figure}[htbp]
\centering \leavevmode
\epsfxsize=0.9\textwidth \epsfbox{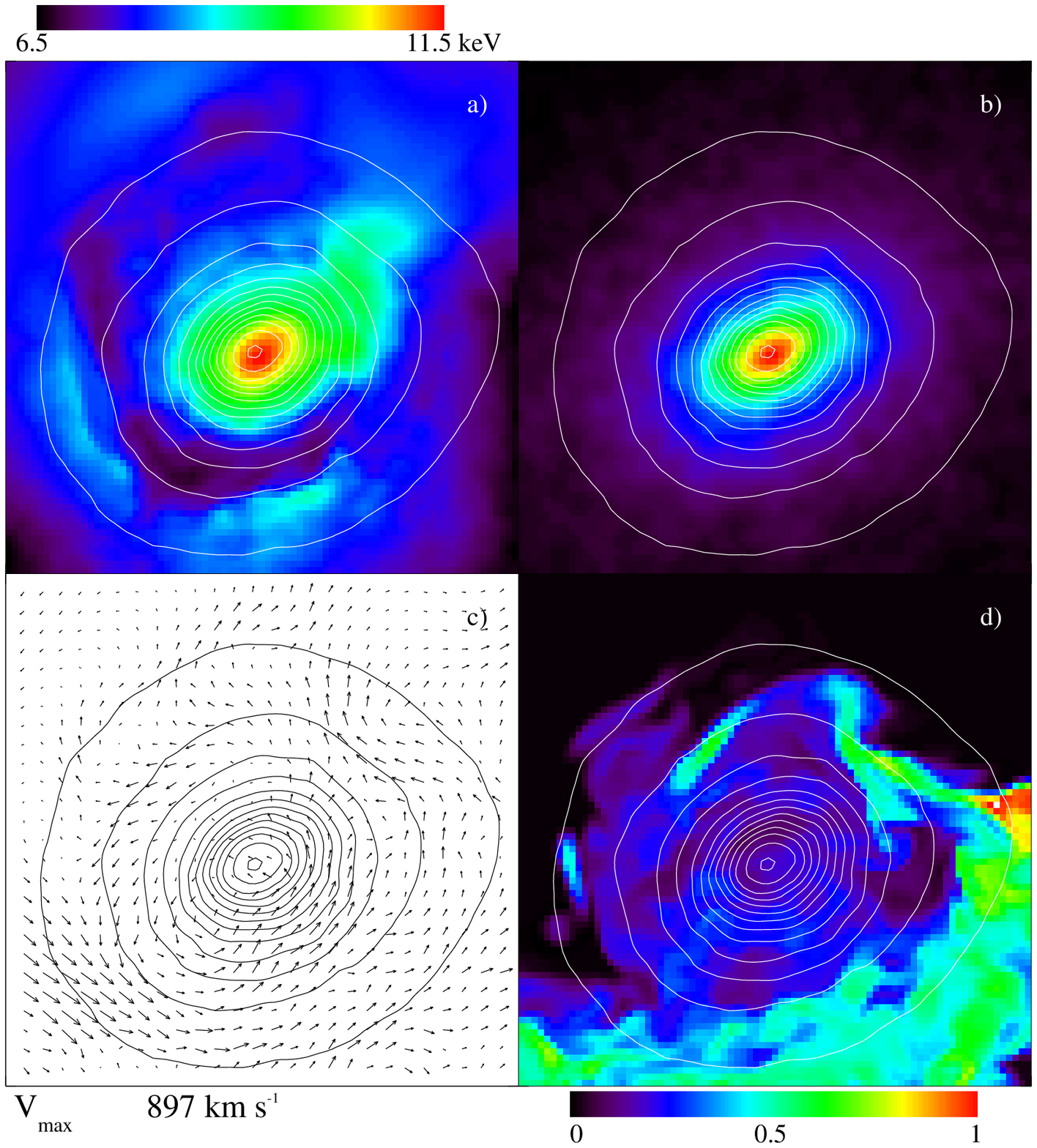}
\caption[]
{ }
\label{simdat6}
\end{figure}

\begin{figure}[htbp]
\centering \leavevmode
\epsfxsize=0.9\textwidth \epsfbox{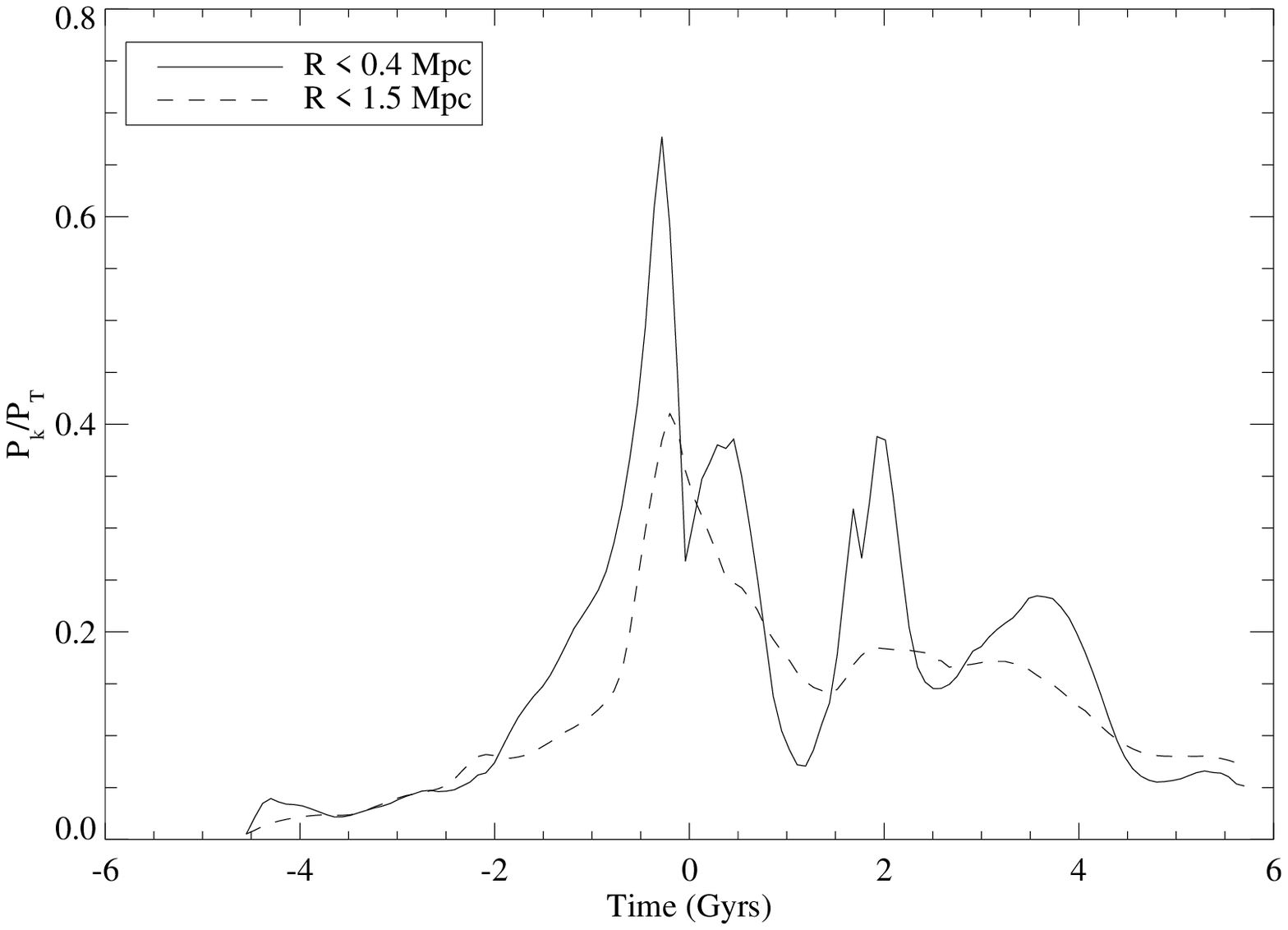}
\caption[]
{ }
\label{pressrat}
\end{figure}

\begin{figure}[htbp]
\centering \leavevmode
\epsfxsize=0.9\textwidth \epsfbox{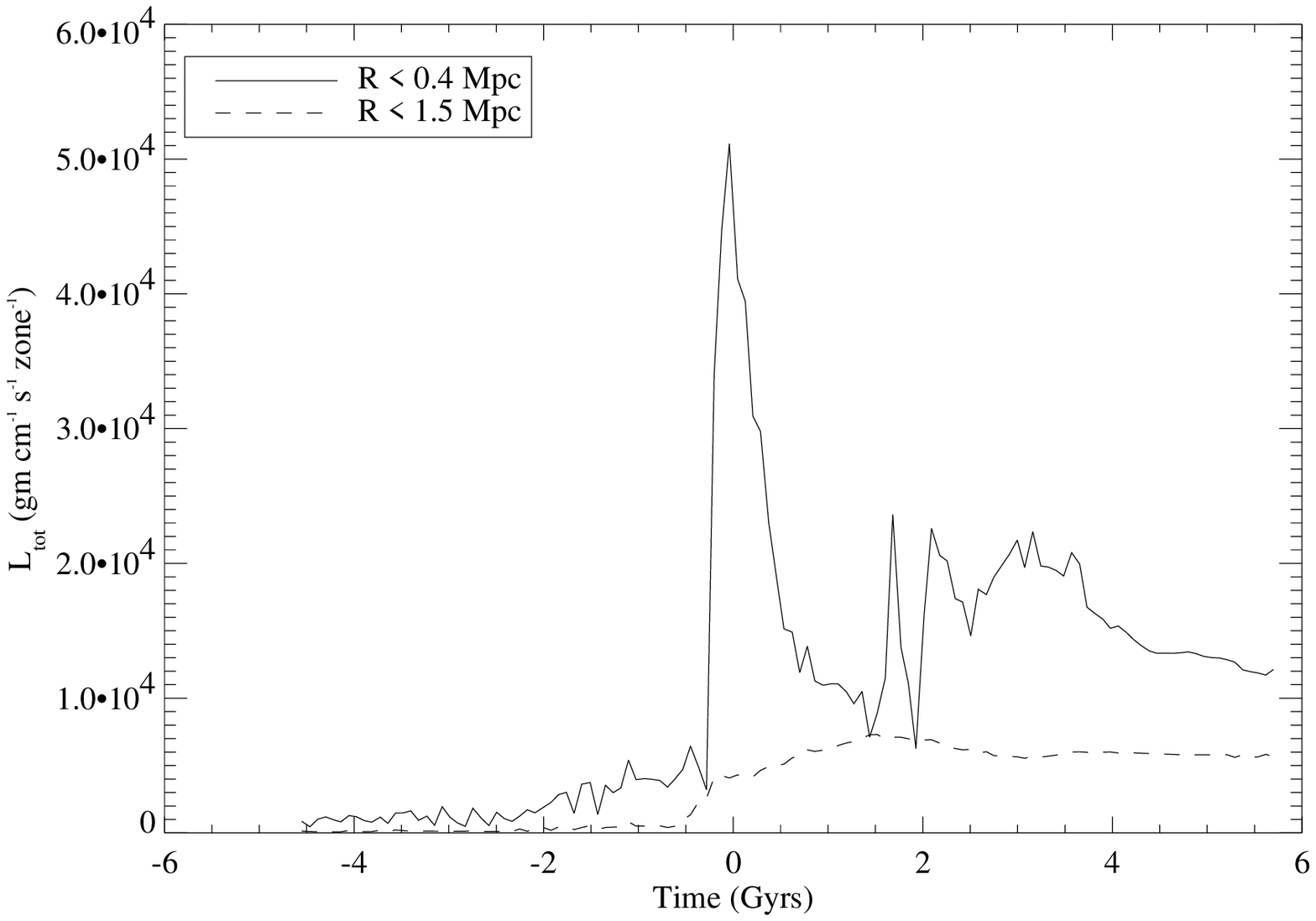}
\caption[]
{ }
\label{angmo}
\end{figure}

\begin{figure}[htbp]
\centering \leavevmode
\epsfxsize=0.9\textwidth \epsfbox{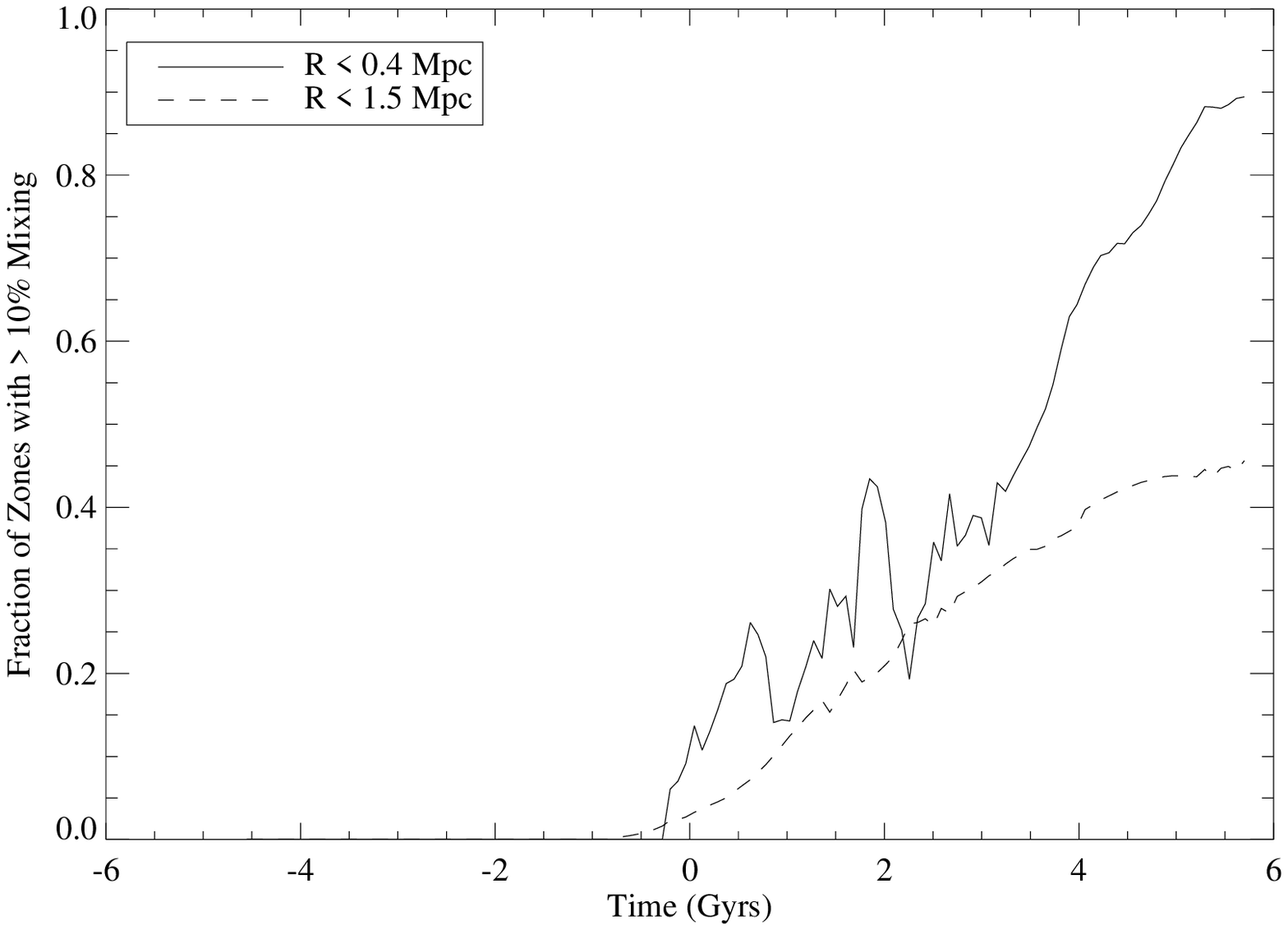}
\caption[]
{ }
\label{color}
\end{figure}

\end{document}